\documentclass[10pt,twocolumn,letterpaper]{article}

\usepackage{wacv}
\usepackage{times}
\usepackage{epsfig}
\usepackage{booktabs}       
\usepackage{graphicx}
\usepackage{amsmath}
\usepackage{amssymb}
\usepackage{subcaption}
\usepackage{xcolor}
\usepackage{enumitem}
\usepackage{bbold}
\usepackage{multirow}
\usepackage{tablefootnote}
\usepackage[accsupp]{axessibility}  

\def\wacvPaperID{974} 

\wacvfinalcopy 

\ifwacvfinal
\fi

\ifwacvfinal
\usepackage[breaklinks=true,bookmarks=false]{hyperref}
\else
\usepackage[pagebackref=true,breaklinks=true,colorlinks,bookmarks=false]{hyperref}
\fi

\begin{document}

\title{PhotoWCT$^2$: Compact Autoencoder for Photorealistic Style Transfer Resulting from Blockwise Training and Skip Connections of High-Frequency Residuals}

\author{Tai-Yin Chiu\\
The University of Texas at Austin
\and
Danna Gurari\\
University of Colorado Boulder
}

\maketitle
\thispagestyle{empty}
\begin{abstract}
   Photorealistic style transfer is an image editing task with the goal to modify an image to match the style of another image while ensuring the result looks like a real photograph.  A limitation of existing models is that they have many parameters, which in turn prevents their use for larger image resolutions and leads to slower run-times.  We introduce two mechanisms that enable our design of a more compact model that we call PhotoWCT$^2$, which preserves state-of-art stylization strength and photorealism.  First, we introduce blockwise training to perform coarse-to-fine feature transformations that enable state-of-art stylization strength in a single autoencoder in place of the inefficient cascade of four autoencoders used in PhotoWCT. Second, we introduce skip connections of high-frequency residuals in order to preserve image quality when applying the sequential coarse-to-fine feature transformations.  Our PhotoWCT$^2$ model requires fewer parameters (e.g., 30.3\% fewer) while supporting higher resolution images (e.g., 4K) and achieving faster stylization than existing models.  
\end{abstract}

\section{Introduction}
\begin{figure}[t!]
    \centering
    \includegraphics[width=\linewidth]{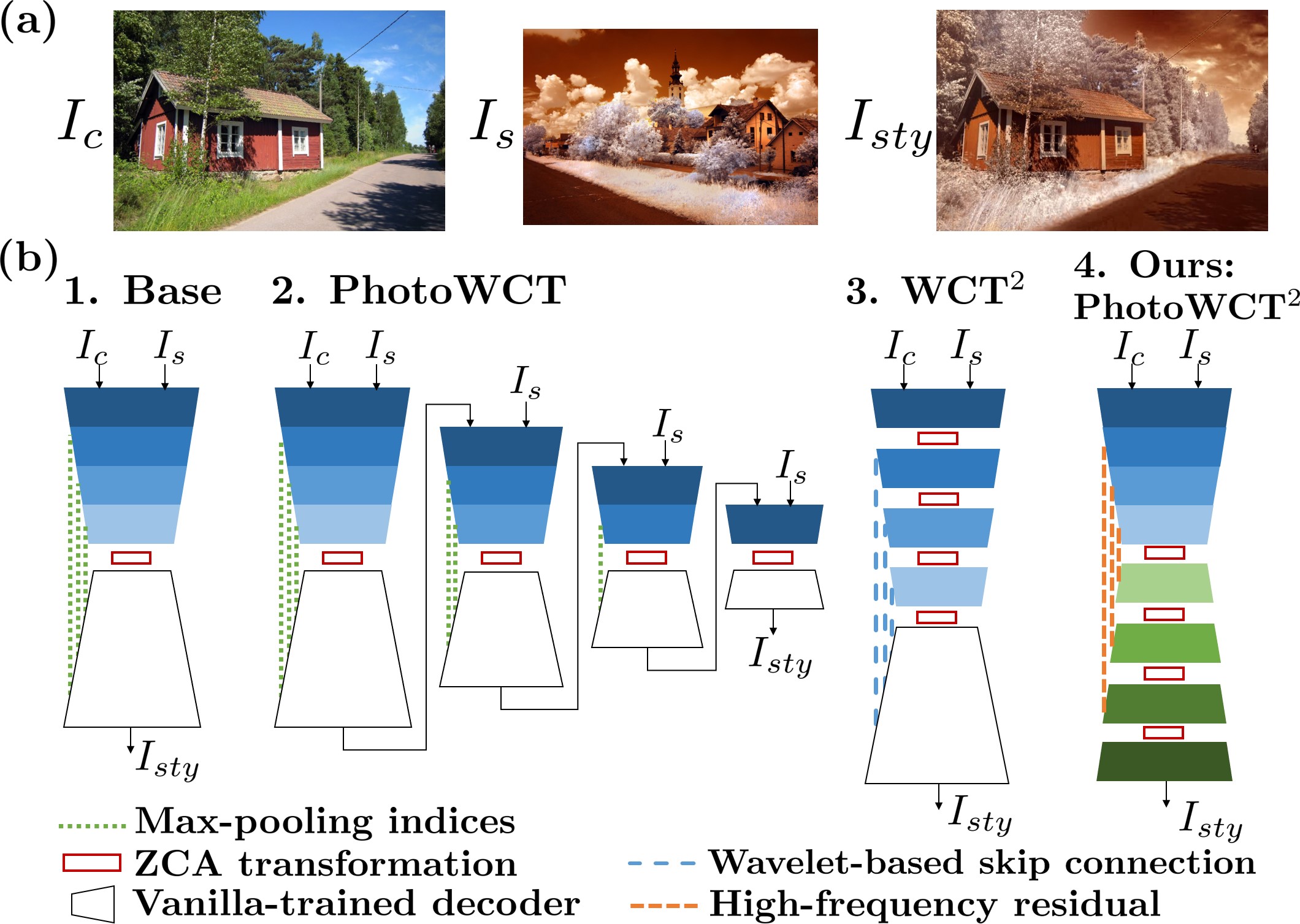}
    \caption{(a) A stylization example using our PhotoWCT$^2$. (b) Comparison of methods for photorealistic style transfer:  (b1) The base framework is an autoencoder that transforms the input content image, $I_c$, and style image, $I_s$, into a stylized image, $I_{sty}$, with a single feature transformation. (b2,b3,b4) The other models perform multi-scale feature transformation with (b2) PhotoWCT~\cite{li2018closed} using four cascaded autoencoders to achieve coarse-to-fine feature transformation, (b3) WCT$^2$~\cite{yoo2019photorealistic} performing fine-to-coarse feature transformation in a single autoencoder using wavelet-based skip connections, and (b4) our PhotoWCT$^2$ realizing coarse-to-fine feature transformation in a single autoencoder with skip connections of high-frequency residuals. }
    \label{fig:overview}
\end{figure}

Photorealistic style transfer is the task of rendering an image in the style of another image such that the result appears like a real photograph to end users (Figure~\ref{fig:overview}a).  A limitation of existing methods is that they are \emph{parameter-heavy}, which results in a number of practical limitations.  First, they cannot support images of 4K resolution (i.e., 8.3 megapixels) and above, also referred to as ultra high definition (UHD) media.  Yet, advancements in technology have led UHD to become standard in commercial products, as exemplified by the increasing number of self-made UHD images and videos shared on online image stocks~\cite{Pexel4k} and YouTube as well as the trend for more movies and TV series on streaming platforms (e.g., Netflix~\cite{Netflix4k}, Amazon Prime Video~\cite{Amazon4k}) to support UHD resolution.  Other practical concerns include the ability to run methods on memory-constrained or power-constrained devices and to support fast stylization.  We aim to introduce a more compact model to address these practical limitations.

Our work, like state-of-art methods~\cite{li2018closed,yoo2019photorealistic}, builds upon the predominant framework for style transfer methods: an autoencoder~\cite{chen2016fast,huang2017arbitrary,li2017universal,li2019learning,sheng2018avatar}. As exemplified in Figure~\ref{fig:overview}b:1, this framework takes as input both a content image and style image ($I_C$, $I_S$), encodes each into a feature representation using a pre-trained network, applies a transformation to alter the content feature with respect to the style feature, and then finally decodes the resulting feature into the stylized image ($I_{sty}$).  Our work centers on redesigning two parameter-heavy mechanisms employed by state-of-art methods into lighter weight representations. 

Our first aim is to redesign the mechanism employed to strongly reflect the new style in the rendered image (i.e., to achieve strong stylization strength).  This mechanism entails using multiple transformations of features of different scales.  The state-of-art method, PhotoWCT~\cite{li2018closed}, employs four cascaded autoencoders to transform the content image by recursively modifying its coarse feature to its fine feature with respect to the corresponding coarse to fine style features (illustrated in Figure~\ref{fig:overview}b:2).  Intuitively, this strengthens the transferred style by first modifying the big picture of the content image with respect to the style image and then gradually fine-tuning its fine-grained details.  However, PhotoWCT's use of multiple autoencoders makes it computationally expensive.  In contrast, WCT$^2$~\cite{li2017universal} uses a single autoencoder to progressively transform the content image.  However, it modifies the content image from its fine feature to its coarse feature with respect to the corresponding fine to coarse style features before decoding the transformed feature into the stylized image (illustrated in Figure~\ref{fig:overview}b:3).  This fine-to-coarse feature transformation is shown experimentally~\cite{li2017universal} to result in weaker stylization strength than the coarse-to-fine feature transformation performed by PhotoWCT.  Intuitively, this worse performance may be because initial fine-tuned details might get overshadowed by later big-picture modifications.  We introduce a redesign that simultaneously embeds the strengths of PhotoWCT and WCT$^2$ while overcoming their limitations.  We achieve this by introducing a novel technique, called blockwise training, that makes it possible to convert the PhotoWCT cascade into a \emph{single compact autoencoder} that performs \emph{coarse-to-fine feature stylization}.

Our second aim is to redesign the mechanism used to recover content information that gets lost by the autoencoder when rendering the stylized image.  Existing methods employ skip connections from the autoencoder's encoder to its decoder for this purpose.  For instance, PhotoWCT~\cite{li2018closed} skip-connects from the encoder's max-pooling layer the indices of computed maximum values (i.e., max-pooling indices) to the paired decoder's max-unpooling layer (illustrated in Figure~\ref{fig:overview}b:2). However, theoretically, max-pooling is lossy~\cite{ye2018deep,yin2017tale} and so these max-pooling indices are not guaranteed to be sufficient for good image reconstruction, which in turn results in content distortion in stylized results.  In contrast, WCT$^2$~\cite{yoo2019photorealistic} introduces skip connections based on wavelets (illustrated in Figure~\ref{fig:overview}b:3) that are guaranteed with signal processing theories~\cite{ye2018deep,yin2017tale} to yield better image reconstruction performance, and are shown experimentally to do so in Section~\ref{sec:bt_with_skip}.  As will be discussed in Section~\ref{sec:methods}, a key reason behind its advantage is that wavelet-based skip connections helps recover high-frequency information that can easily get lost in the encode-decode process. A limitation of wavelet-based skip connections though is that they require many parameters. We introduce an architecture that we call \emph{skip connections of high-frequency residuals} that makes it possible to achieve the advantage of WCT$^2$'s wavelet-based skip connections for better image reconstruction with considerably fewer parameters.

To summarize our key contributions, we introduce a new photorealistic style transfer model, which we call PhotoWCT$^2$ (illustrated in Figure~\ref{fig:overview}b:4), alongside two new mechanisms used to create this model. The first mechanism is \emph{blockwise training} for redesigning the coarse-to-fine feature transformations in PhotoWCT's cascade of autoencoders into a single decoder. The second mechanism is \emph{skip connections of high-frequency residuals} that serves as a lightweight representation of wavelet-based skip connections and enables the success of our blockwise training.  Experiments show our model preserves state-of-art stylization strength and photorealism while achieving a 30.3\% and 15.6\% parameter reduction compared to PhotoWCT and WCT$^2$ respectively.  Moreover, experiments show it can support higher resolution images (i.e., UHD) and achieve faster stylization than existing methods~\cite{li2018closed,yoo2019photorealistic,an2020ultrafast}. Ablation studies demonstrate that existing mechanisms for training and skip connections are insufficient to produce our compact model and so underscore the critical need of our two new mechanisms, blockwise training and skip connections of high-frequency residuals.

\section{Related works}
\label{sec:related_works}
\paragraph{Photorealistic style transfer.}
In 2017, the seminal neural network-based method for photorealistic style transfer was introduced~\cite{luan2017deep}.  To address it is relatively slow due to its need for many iterations of forward passing and backpropagation, new methods~\cite{li2018closed,yoo2019photorealistic,an2020ultrafast,xia2020joint} incurred speed gains by using one forward pass.  Among them, PhotoWCT~\cite{li2018closed} achieves the strongest stylization strength at the expense of a parameter-heavy architecture of four autoencoders.  WCT$^2$~\cite{yoo2019photorealistic} and PhotoNAS~\cite{an2020ultrafast}, in contrast, offer single autoencoder architectures, with WCT$^2$ being superior due to its use of fewer layers and wavelet-based skip connections (which support theory-backed image reconstruction). We introduce a model that achieves comparable stylization strength to the state-of-art PhotoWCT while realizing further advantages over existing models~\cite{li2018closed,yoo2019photorealistic,an2020ultrafast}\footnote{We cannot compare to \cite{xia2020joint} because the code has not been released.  With that said, we expect poorer performance from it because it downsizes images to support high resolution (4K) images and so discards information.}, by requiring fewer parameters, supporting stylization of larger images, and providing faster stylization.

\vspace{-1.1em}\paragraph{Greedy layerwise training.} 
Like traditional greedy layerwise training for autoencoders~\cite{bengio2007greedy,larochelle2009exploring,ngiam2011multimodal,belilovsky2019greedy}, our blockwise training entails splitting an autoencoder into a sequence of sub-model pairs and then training the pairs sequentially.  However, the traditional approach pairs an encoder layer with a decoder layer while our approach pairs an encoder block with a decoder block.  In addition, they are used for different purposes: while the traditional approach centers on \emph{learning an encoder} that represents a specific dataset~\cite{chen2017deep,farahnakian2018deep,pathirage2018structural,liang2018remote,sagheer2019unsupervised} and so uses the decoder as a disposable accessory needed to achieve this aim, our approach instead \emph{fixes a pretrained encoder} during training (e.g., VGG~\cite{simonyan2014very}) in order to learn a decoder that can reproduce the coarse-to-fine features.  Our experiments demonstrate that blockwise training is non-trivial with existing neural network architecture components, failing to produce an effective compact autoencoder (Sections~\ref{sec:training_strategies} and \ref{sec:bt_with_skip}). We introduce skip connections of high-frequency residuals and demonstrate that it overcomes this limitation, enabling the effective use of blockwise training to develop a compact autoencoder for photorealistic style transfer.

\vspace{-1.1em}\paragraph{Skip connections.}
A challenge is how to employ skip connections~\cite{ronneberger2015u,he2016deep} within autoencoders for photorealistic style transfer.  In particular, when an autoencoder consists of an encoder that is a fixed pre-trained model and a decoder that learns its inverse function, directly connecting an output from a layer $l_e$ in the encoder to some layer $l_d$ of the decoder results in a short circuit phenomenon~\cite{an2020ultrafast}.  This means the connection is so informative that it overshadows the middle layers between $l_e$ and $l_d$, such that the middle layers will not affect the pixel values in the output of the decoder after training.  Numerous variants of skip connections address this issue, including indices of maximal values between max-pooling/unpooling layers for PhotoWCT~\cite{li2018closed}, instance-normalized skip-connected features for PhotoNAS~\cite{an2020ultrafast}, and a theoretically motivated
wavelet-based approach for WCT$^2$~\cite{yoo2019photorealistic}.  We simplify the wavelet-based architecture into a computationally light variant we call skip connections of high-frequency residuals.

As will be detailed in Section~\ref{sec:sc_residual}, the computation of a high-frequency residual is similar to that of the first difference image in a Laplacian pyramid~\cite{burt1987laplacian}. However, while a Laplacian pyramid is built upon an image for multiple levels, a high-frequency residual is computed from a feature map and does not form a pyramid. Moreover, most previous works~\cite{denton2015deep,ghiasi2016laplacian,lai2017deep,xu2018lapran,fu2019lightweight,anwar2020densely} that integrate a Laplacian pyramid into neural networks heuristically leverage the concept that a Laplacian pyramid preserves high-frequency details from the input image to generate images of better quality. We extend prior work by providing a theoretical explanation why our approach can realize an autoencoder for coarse-to-fine feature transformation for photorealistic style transfer.

\section{Method}
\label{sec:methods}
We now introduce our new model PhotoWCT$^2$ and our two mechanisms that enable its creation: blockwise training and skip connections of high-frequency residuals.

\subsection{Background}
To begin, we describe the parameter-heavy mechanisms used in state-of-art photorealistic style transfer methods that we aim to redesign into compact representations.

\label{sec:background}
\vspace{-1.1em}\paragraph{PhotoWCT's coarse-to-fine feature transformations.} As summarized in the Introduction and illustrated in Figure~\ref{fig:overview}b:2, PhotoWCT~\cite{li2018closed} consists of a cascade of four autoencoders AEC$_N$'s ($N$ = 1, 2, 3, 4), where each includes an encoder $\textit{enc}_N$ and decoder $\textit{dec}_N$.  $\textit{enc}_N$ is a pretrained network, specifically VGGNet, from the input layer to the $\textit{reluN\_1}$ layer. $\textit{dec}_N$ is structurally symmetric to $\textit{enc}_N$.  To realize the coarse-to-fine feature transformation, the cascade of four autoencoders is in the order from $N=4$ to $N=1$.  Specifically, content and style images are first encoded by $\textit{enc}_4$ into the $\textit{relu4\_1}$ features. The $\textit{relu4\_1}$ content feature is then transformed with reference to the $\textit{relu4\_1}$ style feature using a ZCA feature transformation~\cite{li2017universal,chiu2019understanding}. The transformed feature is then decoded by $\textit{dec}_4$ to become an image $I_4$. The three steps of encoding, transformation, and decoding repeat in the next three rounds of $N=3,2,1$, with $I_{N+1}$ as the content image, until the stylized image $I_1$ is decoded by $\textit{dec}_1$.  Finally, image smoothing (using guided filtering) is applied as a post-processing step to $I_1$ to remove undesired artifacts in the final stylized image.\footnote{The original code for this step has a bug.  We describe this issue and our fix in the Supplementary Materials.}  

\begin{figure}[t!]
    \centering
    \includegraphics[width=\linewidth]{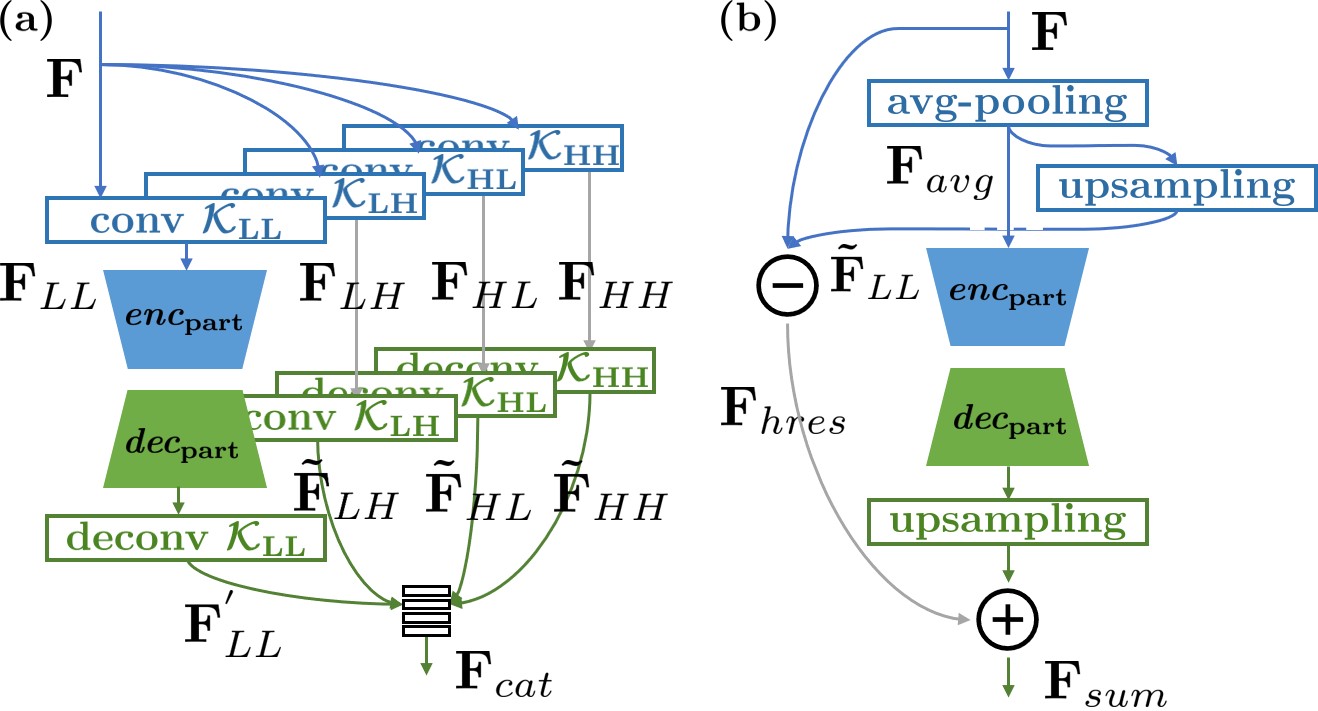}
    \vspace{-1.75em}
    \caption{Shown are the middle layers of an autoencoder between a pooling and an unpooling layer for (a) WCT$^2$\cite{yoo2019photorealistic} and (b) our method. (a) The wavelet-based skip connection uses the wavelet pooling (blue rectangles) and wavelet unpooling (green rectangles) to improve the image reconstruction ability. (b) Our skip connection of the high-frequency residual simplifies this wavelet-based skip connection into a more compact representation. }
    \label{fig:skip_connection}
    \vspace{-0.25em}
\end{figure}

\vspace{-1.1em}\paragraph{Wavelet-based skip connections.}
The architecture of wavelet-based skip connections, which were introduced as part of WCT$^2$~\cite{yoo2019photorealistic}, is shown in Figure~\ref{fig:skip_connection}a.  Note that WCT$^2$ is an autoencoder which emulates AEC$_4$ in PhotoWCT while replacing its max-pooling/unpooling layers with wavelet pooling/unpooling layers for better image reconstruction. As exemplified in Figure~\ref{fig:skip_connection}a, its wavelet pooling layer, when given a feature $\mathbf{F}$, produces a low-frequency component $\mathbf{F}_{LL}$ and three high-frequency components $\mathbf{F}_{LH}$, $\mathbf{F}_{HL}$, and $\mathbf{F}_{HH}$.  Structurally, $\mathbf{F}_{LL}$ propagates through the middle layers ($\textit{enc}_{part}$-$\textit{dec}_{part}$ $\triangleq$ AEC$_{part}$) of the network.  The skip connections $\mathbf{F}_{LH}$, $\mathbf{F}_{HL}$, $\mathbf{F}_{HH}$ are then aggregated with $\mathbf{F}_{LL}$ at the corresponding wavelet unpooling layer in the decoder.  The intuition is that most information of $\mathbf{F}$ gets carried by $\mathbf{F}_{LL}$, and the high-frequency information $\mathbf{F}_{LH}$, $\mathbf{F}_{HL}$, and $\mathbf{F}_{HH}$ can be supplemented to improve the model's reconstruction.

Mathematically, a wavelet pooling/unpooling performs four depthwise convolutions/deconvolutions with stride 2 using the following Haar wavelet kernels:
\begin{equation}
    \mathcal{K}_{LL} = \frac{1}{2}
    \begin{bmatrix}
        1 &\hspace{-0.5em} 1\\
        1 &\hspace{-0.5em} 1
    \end{bmatrix},
    \mathcal{K}_{LH} = \frac{1}{2}
    \begin{bmatrix}
        -1 &\hspace{-0.5em} 1\\
        -1 &\hspace{-0.5em} 1
    \end{bmatrix},
    \mathcal{K}_{HH} = \frac{1}{2}
    \begin{bmatrix}
        1 &\hspace{-0.5em} -1\\
        -1 &\hspace{-0.51em} 1
    \end{bmatrix}
    \label{eq:wavelet_kernels}
\end{equation}
and $\mathcal{K}_{HL}$ = $\mathcal{K}_{LH}^\mathrm{T}$. Let $(\mathcal{K}_{ij} * \mathbf{f})_{\downarrow 2}$ and $(\mathcal{K}_{ij} * \mathbf{f})_{\uparrow 2}$ denote the 2-strided convolution and deconvolution of $\mathcal{K}_{ij}$ ($i,j$ $\in$ $\{L,H\}$) and a feature $\mathbf{f}$, respectively. 

The wavelet unpooling output $\mathbf{F}_{cat}$ in Figure~\ref{fig:skip_connection}a is the concatenation of four components $\mathbf{F}^{'}_{LL}$, $\mathbf{\tilde{F}}_{LH}$, $\mathbf{\tilde{F}}_{HL}$, and $\mathbf{\tilde{F}}_{HH}$, mathematically described as:
\begin{equation}
    \mathbf{F}^{'}_{LL} = (\mathcal{K}_{LL} * \text{AEC}_{part}((\mathcal{K}_{LL} * \mathbf{F})_{\downarrow 2})))_{\uparrow 2},
    \label{eq:F_ll_prime}
\end{equation}
\begin{equation}
    \mathbf{\tilde{F}}_{ij} = (\mathcal{K}_{ij} * (\mathcal{K}_{ij} * \mathbf{F})_{\downarrow 2}))_{\uparrow 2},~i,j \in \{L,H\}.
    \label{eq:F_tilde_ij}
\end{equation}
\noindent
While this type of skip connection prevents the loss of high-frequency information of the input image and so leads to better image reconstruction~\cite{yoo2019photorealistic}, it is computationally expensive.  For comparison, it requires four times as many parameters as the max-pooling indices skip connection used by PhotoWCT, since the wavelet unpooling output $\mathbf{F}_{cat}$ has four times the channel length of PhotoWCT's max-unpooling output.

\subsection{Our approach: PhotoWCT$^2$}
\label{sec:model_redesign}

\subsubsection{Model architecture}
\label{sec:training_setting}
\begin{figure}[t!]
    \centering
    \includegraphics[width=\linewidth]{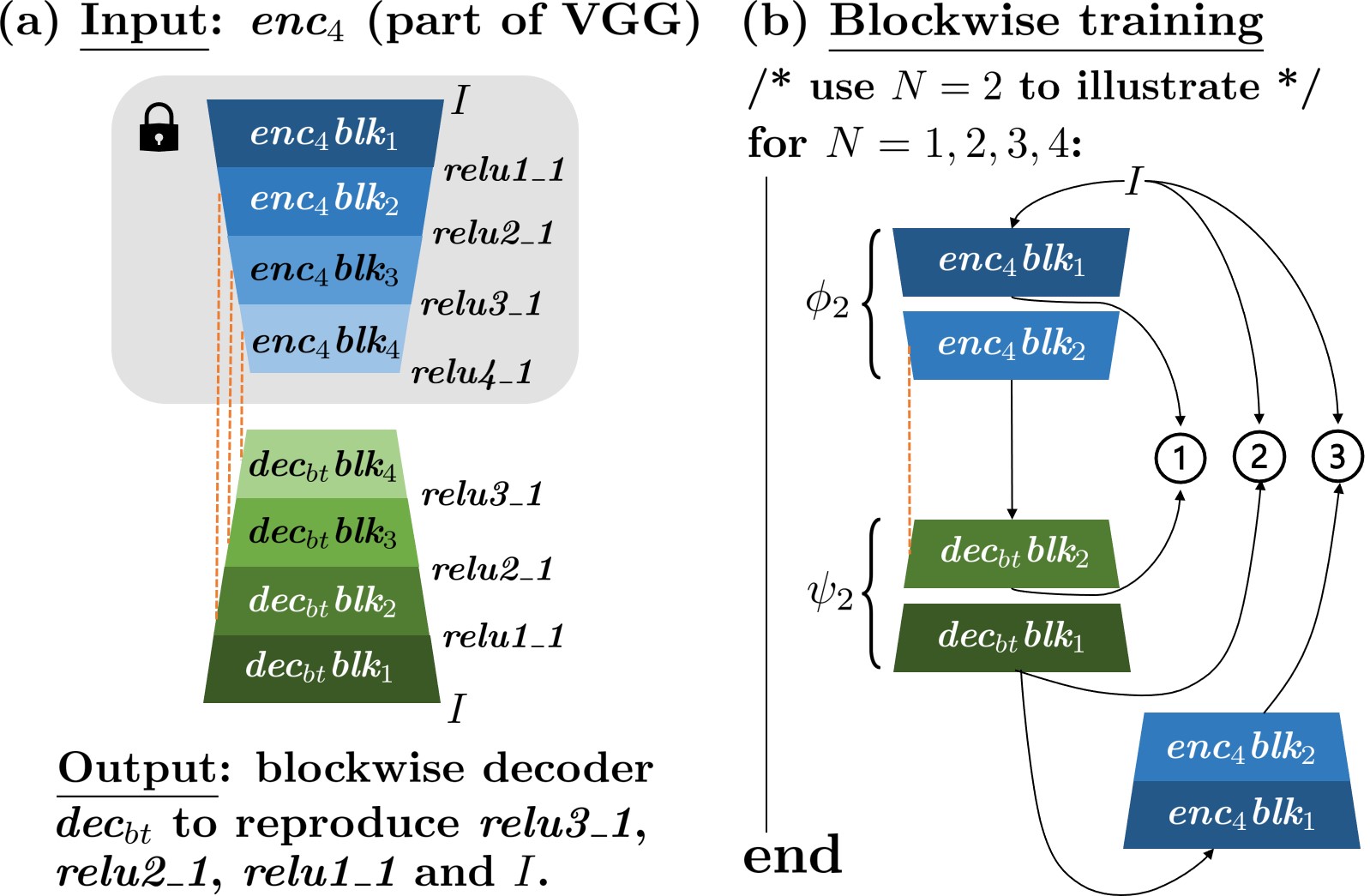}
    \caption{(a) Overview of our PhotoWCT$^2$ model architecture and (b) illustration of our blockwise training strategy needed to effectively support coarse-to-fine feature transformation in a single autoencoder.  The three shown circles indicate the losses blockwise training minimizes, with 1 representing function inversion loss, 2 representing image reconstruction loss, and 3 representing perceptual loss.}
    \label{fig:blockwise_training}
\end{figure}

We design our model as an autoencoder AEC$_{bt}$.  An overview of its architecture is shown in Figure~\ref{fig:blockwise_training}a.  

For the encoder, we rely on PhotoWCT's encoder $\textit{enc}_4$ (described in Section~\ref{sec:background}).   As exemplified in Figure~\ref{fig:blockwise_training}a, $\textit{enc}_4$ is split into the series of blocks $\textit{enc}_4\textit{blk}_1$, $\textit{enc}_4\textit{blk}_2$, $\textit{enc}_4\textit{blk}_3$, and $\textit{enc}_4\textit{blk}_4$.  The output layer of the block $\textit{enc}_4\textit{blk}_N$ is the $\textit{reluN\_1}$ layer in VGGNet.

We design the decoder, which we call $\textit{dec}_{bt}$, to be structurally symmetric to $\textit{enc}_4$.  As exemplified in Figure~\ref{fig:blockwise_training}a, $\textit{dec}_{bt}$ is split into the series of blocks $\textit{dec}_{bt}\textit{blk}_4$, $\textit{dec}_{bt}\textit{blk}_3$, $\textit{dec}_{bt}\textit{blk}_2$, and $\textit{dec}_{bt}\textit{blk}_1$.  
We design the $\textit{dec}_{bt}\textit{blk}_N$'s to be structurally symmetric to the $\textit{enc}_4\textit{blk}_N$'s, with the goal that it will learn the inverse function of $\textit{enc}_4\textit{blk}_N$, i.e., to convert $\textit{reluN\_1}$ features to $\textit{relu(N-1)\_1}$ features. As such, after training, the decoder taking the $\textit{relu4\_1}$ feature at the bottleneck should be able to sequentially reproduce the $\textit{relu3\_1}$, $\textit{relu2\_1}$, and $\textit{relu1\_1}$ features and the input image. 

In order to realize the coarse-to-fine feature transformations, we embed feature transformations at the bottleneck between the encoder and decoder as well as the outputs of $\textit{dec}_{bt}\textit{blk}_4$, $\textit{dec}_{bt}\textit{blk}_3$, and $\textit{dec}_{bt}\textit{blk}_2$ blocks (illustrated in Figure~\ref{fig:overview}b:4).  Recall that the purpose of each transformation is to alter each content feature with respect to each style feature at a different scale.  Following PhotoWCT, we employ ZCA transformations.  We will describe in Section~\ref{sec:blockwise-training} blockwise training, which is the critical ingredient to make this compact design possible.

We also integrate skip connections into the autoencoder in order to improve not only the image reconstruction, as shown for WCT$^2$~\cite{yoo2019photorealistic}, but also the feature reconstruction in our blockwise training (as will be shown in Section~\ref{sec:bt_with_skip}).  We insert our new skip connections of high-frequency residuals in the same positions as used for PhotoWCT's max-pooling 
indices skip connections.  Consequently, in $\textit{enc}_4\textit{blk}_N$ for $N=2,3,4$, we replace the original max-pooling with an average-pooling layer from which a skip connection of the high-frequency residual links to the counterpart upsampling layer in $\textit{dec}_{bt}\textit{blk}_N$.  Our new, compact skip connection design will be described in Section~\ref{sec:sc_residual}.

Finally, as done for PhotoWCT, image smoothing (via guided filtering) is applied as a post-processing step. 

\subsubsection{Blockwise training}
\label{sec:blockwise-training}
We propose two methods: end-to-end training and blockwise training to realize the four function inversions for the decoder. End-to-end training enables the decoder to learn the four function inversions at once. While end-to-end training is good enough to invert the functions, we improve upon it by proposing blockwise training. Our blockwise training approach is illustrated in Figure~\ref{fig:blockwise_training}b.  As shown, learning of the four function inversions for the decoder is distributed into four steps such that, at each step, a decoder block $\textit{dec}_{bt}\textit{blk}_N$ learns the inverse function of $\textit{enc}_4\textit{blk}_N$.    It will be demonstrated in Section~\ref{sec:training_strategies} that this training approach enables more faithful feature reconstruction and image reconstruction in the decoder than alternatives including end-to-end training.  
 
 Mathematically, the decoder blocks $\textit{dec}_{bt}\textit{blk}_N$'s are trained in the order from $N=1$ to $N=4$ by minimizing the loss $\mathcal{L}_N$:
\begin{equation}
    \begin{aligned}
    \mathcal{L}_N(I) = &~\mathbb{1}_{N\neq1}||\phi_{N-1}(I)-\textit{dec}_{bt}\textit{blk}_N(\phi_N(I))||^2_2 \\
    & + ||I-\psi_N(\phi_N(I))||^2_2 \\
    & + \lambda||\phi_N(I)-\phi_N(\psi_N(\phi_N(I)))||^2_2,
    \end{aligned}
    \label{eq:loss_inward}
\end{equation}
where $\phi_N$ and $\psi_N$ are the functions of the series $\{\textit{enc}_4\textit{blk}_1, \dots, \textit{enc}_4\textit{blk}_N\}$ and $\{\textit{dec}_{bt}\textit{blk}_N, \dots, \textit{dec}_{bt}\textit{blk}_1\}$, respectively,  $\mathbb{1}_{N\neq1}$ is an indicator function equal to one when $N\neq 1$ and zero when $N = 1$, and $\lambda$ is set to one for $N\neq 1$ and zero for $N = 1$. The three terms in Equation~\ref{eq:loss_inward} are the function inversion, image reconstruction, and perceptual losses, respectively. When training a decoder block, the previously trained blocks and the encoder are fixed.\footnote{Due to space constraints, we show in the Supplementary Materials two advantages of training from $N=1$ to $N=4$ over the reversed order ($N=4$ to $N=1$): better image reconstruction and ease of redesigning a cascade of fewer autoencoders (e.g., 3) into a single autoencoder.}

\vspace{-2mm}
\subsubsection{Skip connections of high-frequency residuals}
\label{sec:sc_residual}

Our skip connection of a high-frequency residual is illustrated in Figure~\ref{fig:skip_connection}b.  It helps achieve the aim of end-to-end and blockwise trainings of feature/image reconstruction, utilizing and simplifying wavelet-based skip connections into a less computationally expensive representation by replacing the Haar convolutions with average pooling, upsampling and substraction and the Haar deconvolutions with upsampling and addition (as observed when comparing Figure~\ref{fig:skip_connection}a to Figure~\ref{fig:skip_connection}b).  In doing so, it redesigns the concatenation of $\mathbf{F}^{'}_{LL}$, $\mathbf{\tilde{F}}_{LH}$, $\mathbf{\tilde{F}}_{HL}$, and $\mathbf{\tilde{F}}_{HH}$ for wavelet-based skip connections into an addition for our approach, thereby enabling the channel length of our outcome $\mathbf{F}_{sum}$ to become one fourth that of $\mathbf{F}_{cat}$ from the wavelet-based skip connection.  

Our motivation for this addition-based approach is an approximation resulting from the observation that the low-frequency parts of an image are much better reconstructed by an autoencoder than the high-frequency edges, as exemplified in Figure~\ref{fig:reconstruction}b. Taking advantage of this observation, we assume a low-frequency feature $\mathbf{f}$ can be approximately reconstructed by AEC$_{part}$ (described in Section~\ref{sec:background}), i.e., AEC$_{part}(\mathbf{f})$ $\approx$ $\mathbf{f}$. This assumption implies the following approximation: 
\begin{equation}
    \begin{aligned}
        \mathbf{F}^{'}_{LL} &= (\mathcal{K}_{LL} * \text{AEC}_{part}((\mathcal{K}_{LL} * \mathbf{F})_{\downarrow 2})))_{\uparrow 2} \\
        &\approx (\mathcal{K}_{LL} * (\mathcal{K}_{LL} * \mathbf{F})_{\downarrow 2}))_{\uparrow 2} = \mathbf{\tilde{F}}_{LL}.
    \end{aligned}
    \label{eq:f_ll_prime_approx}
\end{equation}
With the Haar wavelet expansion $\mathbf{F} = \mathbf{\tilde{F}}_{LL} + \mathbf{\tilde{F}}_{LH} + \mathbf{\tilde{F}}_{HL} + \mathbf{\tilde{F}}_{HH}$ and the substitution $\mathbf{\tilde{F}}_{LL}$ $\approx$ $\mathbf{F}^{'}_{LL}$,
we arrive at the following approximation $\mathbf{F}^{'}_{LL}$ + $\mathbf{\tilde{F}}_{LH}$ + $\mathbf{\tilde{F}}_{HL}$ + $\mathbf{\tilde{F}}_{HH}$ $\approx$ $\mathbf{F}$. This implies that with addition as feature aggregation, a wavelet pooling input $\mathbf{F}$ in the encoder is likely to be reconstructed at the wavelet unpooling layer in the decoder.  

In describing the implementation of the \emph{encoder} part in Figure~\ref{fig:skip_connection}b, let $\mathbf{F}_{hres}$ be the sum of high-frequency components $\mathbf{\tilde{F}}_{LH}$ $+$ $\mathbf{\tilde{F}}_{HL}$ $+$ $\mathbf{\tilde{F}}_{HH}$. We call $\mathbf{F}_{hres}$ the \textit{high-frequency residual} of $\mathbf{F}$ since it is the difference between $\mathbf{F}$ and the low-frequency component $\mathbf{\tilde{F}}_{LL}$. By substituting $\mathcal{K}_{LL}$ in Equation~\ref{eq:F_tilde_ij} with its matrix form, $\mathbf{\tilde{F}}_{LL}$ can be rewritten as follows:
\begin{equation}
   \begin{aligned}
   \mathbf{\tilde{F}}_{LL} &= \Big( \frac{1}{2}
    \begin{bmatrix}
        1 &\hspace{-0.5em} 1\\
        1 &\hspace{-0.5em} 1
    \end{bmatrix} * \Big( \frac{1}{2}
    \begin{bmatrix}
        1 &\hspace{-0.5em} 1\\
        1 &\hspace{-0.5em} 1
    \end{bmatrix} * \mathbf{F} \Big)_{\downarrow 2}\Big)_{\uparrow 2}
    \\
    &= \Big(
    \begin{bmatrix}
        1 &\hspace{-0.5em} 1\\
        1 &\hspace{-0.5em} 1
    \end{bmatrix} * \Big( \frac{1}{4}
    \begin{bmatrix}
        1 &\hspace{-0.5em} 1\\
        1 &\hspace{-0.5em} 1
    \end{bmatrix} * \mathbf{F} \Big)_{\downarrow 2}\Big)_{\uparrow 2}
    \\
    &= \mathtt{upsampling}(\mathtt{avgpooling}(\mathbf{F})).
    \end{aligned}
    \label{eq:f_ll_rewritten}
\end{equation}
Therefore, the computation of $\mathbf{F}_{hres}$ simplifies to the following: $\mathbf{F}$ is first average-pooled to become a low-frequency feature $\mathbf{F}_{avg}$ $=$ $\mathtt{avgpooling}(\mathbf{F})$, which in turn is upsampled and subtracted from $\mathbf{F}$. Note that if we generalize to build a pyramid on a feature map but not an image and replace the 2-by-2 matrices of ones in Equation~\ref{eq:f_ll_rewritten} by 5-by-5 Gaussian matrices, $\mathbf{F}_{hres}$ becomes the first difference `image' in the Laplacian pyramid~\cite{burt1987laplacian} of $\mathbf{F}$. Also note that different from the Haar pyramid~\cite{adelson1987orthogonal} of $\mathbf{F}$, where the first level saves $\mathbf{F}_{ij}$ ($i,j$ $\in$ $\{L,H\}$) of the half size of $\mathbf{F}$, our framework saves $\mathbf{F}_{LL}$ and $\mathbf{F}_{hres}$ of the same size as $\mathbf{F}$.

To reproduce the feature $\mathbf{F}$ in the \emph{decoder} by skip-connecting $\mathbf{F}_{hres}$, $\mathbf{F}_{avg}$ is forward-passed through AEC$_{part}$, upsampled, and added to $\mathbf{F}_{hres}$ as illustrated in Figure~\ref{fig:skip_connection}b. The sum feature $\mathbf{F}_{sum}$ reproduces $\mathbf{F}$ under the assumption AEC$_{part}(\mathbf{f})$ $\approx$ $\mathbf{f}$ for a low-frequency feature $\mathbf{f}$:
\begin{equation}
    \begin{aligned}
    \mathbf{F}_{sum} &= \mathtt{upsampling}(\text{AEC}_{part}(\mathbf{F}_{avg})) + \mathbf{F}_{hres}\\
    &\approx \mathtt{upsampling}(\mathbf{F}_{avg}) + \mathbf{F}_{hres} \\
    &= \mathbf{\tilde{F}}_{LL} + (\mathbf{\tilde{F}}_{LH} + \mathbf{\tilde{F}}_{HL} + \mathbf{\tilde{F}}_{HH}) = \mathbf{F}.
    \end{aligned}
\end{equation}

\section{Experiments}
\label{sec:experiments}
\setlength{\tabcolsep}{3pt}
\begin{table*}[!ht]
    \centering\small
    \begin{tabular}{l c c c c c c c c c c c c }
    \toprule
        \multirow{3}{*}{Model} & \multicolumn{2}{c}{(a) Size} & \multicolumn{5}{c}{(b) Speed performance} & \multicolumn{5}{c}{(c) Image quality \& Stylization strength} \\
    \cmidrule[0.5pt](lr){2-3}\cmidrule[0.5pt](lr){4-8}\cmidrule[0.5pt](lr){9-13}
         & \multirow{2}{*}{\# par} & \multirow{2}{*}{\# layer} & \multirow{2}{*}{1024$\times$512} & HD & FHD & QHD & 4K & BRIS & NIQE & NIMA-q & NIMA-a & \multirow{2}{*}{$\bar{\mathcal{L}}_{s,m}$} \\
         & & & &1280$\times$720 & 1920$\times$1080 & 2560$\times$1440 & 3840$\times$2160 & (27.4) & (3.19) & (5.11) & (5.27) & \\
    \midrule
    PhNAS & 40.24M & 35 & 0.23 & OOM & OOM & OOM & OOM & 33.0 & 3.24 & 4.75 & 4.92 & 1.02\\
    WCT$^2$ & 10.12M & \bf{24} & 0.30 & 0.43 & 0.80 & OOM & OOM & \bf{30.8} & 3.07 & \bf{4.91} & 5.01 & 0.80\\
    PhWCT & 8.35M & 48 & 0.21+0.03 & 0.32+0.06 & 0.61+0.14 & 1.01+0.23 & OOM & 31.8 & \bf{2.90} & 4.88 & 5.06 & \bf{-0.70} \\
    Ours (E2E) & \multirow{2}{*}{\bf{7.05M}} & \multirow{2}{*}{\bf{24}} & \multirow{2}{*}{\bf{0.18+0.03}}  & \multirow{2}{*}{\bf{0.24+0.06}} & \multirow{2}{*}{\bf{0.39+0.14}} & \multirow{2}{*}{\bf{0.59+0.23}} & \multirow{2}{*}{\bf{1.22+0.54}} & 31.7 & 2.91 & 4.90 & \bf{5.10} & -0.66\\
    Ours (BT) & & & & & & & & 31.6 & \bf{2.90} & 4.90 & \bf{5.10} & -0.69\\
    \bottomrule
    \end{tabular}
    \vspace{-2mm}
    \caption{Characteristics of our models (PhotoWCT$^2$ (E2E) and PhotoWCT$^2$ (BT)) and three baselines PhotoNAS (PhNAS), WCT$^2$, and PhotoWCT (PhWCT) in terms of model size, speed performance, image quality, and stylization performance. Our models are the most lightweight, the fastest, able to handle the largest resolution (4K), while preserving the good image quality and strong stylization strength of existing state-of-art methods. BRIS: BRISQUE. NIMA-q/a: NIMA-quality/aesthetic. OOM: out-of-memory. (The four values in the parentheses for (c) are scores for pristine images.)}
    \label{tab:model_charateristics}
\end{table*}

We now evaluate our style transfer model and two mechanisms for creating it: blockwise training (BT) and skip connections of high-frequency residuals. For comparison, we also include our model resulting from end-to-end (E2E) training. We refer to our two models as PhotoWCT$^2$ when distinction is not necessary, and PhotoWCT$^2$ (E2E) and PhotoWCT$^2$ (BT) otherwise. We conduct all experiments on an Nvidia 1080-Ti GPU with 11GB memory.


\subsection{Model size and speed}
\label{sec:size_speed}
First, we assess model size and two benefits that arise from a more compact model: model  speed and support for higher resolution images. 

\vspace{-1.1em}\paragraph{Baselines.} 
For comparison, we evaluate two top-performing models, PhotoWCT~\cite{li2018closed} and WCT$^2$~\cite{yoo2019photorealistic}, as well as the more recent PhotoNAS model~\cite{an2020ultrafast}.

\vspace{-1.1em}\paragraph{Dataset.}
We test all models on five resolutions: $1024\times512$,  $1280$$\times$$720$ (HD), $1920$$\times$$1080$ (Full HD), $2560$$\times$$1440$ (Quad HD), and $3840$$\times$$2160$ (4K UHD). To efficiently collect images, we download a 4K video~\cite{4kvideo} from YouTube and sample a frame per second to collect 100 frames. We then downsample each frame to the other lower resolutions.

\vspace{-1.1em}\paragraph{Results.}
Table~\ref{tab:model_charateristics}(a,b) shows the number of parameters, the number of layers\footnote{The following types of layers are counted: convolution, deconvolution, max-pooling, average pooling, and upsampling layers.} on the mainstream path of each model  (i.e., skip connections are excluded), and the models' stylization speed for different resolutions. 

As observed in Table~\ref{tab:model_charateristics}a, our PhotoWCT$^2$ model is the most lightweight. It uses $82.5\%$ fewer parameters than PhotoNAS (7.05M vs. 40.24M), $30.3\%$ fewer parameters than WCT$^2$ (7.05M vs. 10.12M), and $15.6\%$ fewer parameters than PhotoWCT (7.05M vs. 8.35M).  

PhotoWCT$^2$ is also the only model that can handle all the tested resolutions.  This implies that it is the only approach that can stylize UHD (i.e., 4K) images. 

Additionally, our PhotoWCT$^2$ can stylize images at the fastest speeds among all models. For instance, PhotoWCT$^2$ is 0.27 and 0.22 seconds faster than WCT$^2$ and PhotoWCT for FHD images, taking 0.53 seconds.  PhotoWCT$^2$ also saves $0.42$ seconds compared to PhotoWCT (0.82 vs. 1.24) for QHD image stylization. 

\subsection{Image quality and stylization strength}
\label{sec:quality_and_strength}
We next assess to what extent our compressed model can preserve the advantages of existing methods, specifically the ability to generate high quality images and render images with a strong stylization strength.

\vspace{-1.1em}\paragraph{Baselines.}
For comparison, we again evaluate PhotoNAS, PhotoWCT, and WCT$^2$.  As an upper bound, we also evaluate the quality of the original content images, which we refer to as ``pristine" images.  This enables examination of the extent to which stylization downgrades the original quality.

\vspace{-1.1em}\paragraph{Dataset.}
We use the modified version of the DPST dataset~\cite{luan2017deep}.\footnote{The original dataset consists of 60 content-style image pairs including some toy examples, which are excluded here.} We swap the roles of content and style images in each pair to generate more examples, resulting in 100 stylized images per model. 


\begin{figure*}[!t]
    \centering
    \includegraphics[width=0.9\linewidth]{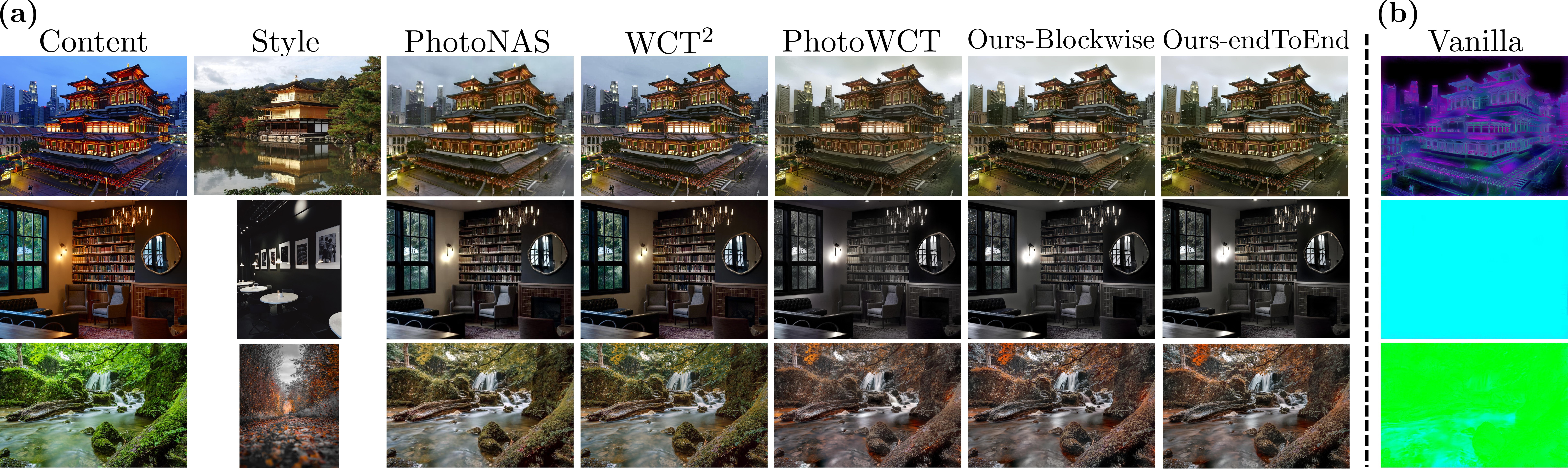}
    \vspace{-2mm}
    \caption{Stylized results from (a) different models and (b) our model trained with vanilla training. Our end-to-end trained model (PhotoWCT$^2$ (E2E)) and blockwisely trained model (PhotoWCT$^2$ (BT)) can produce visually pleasant results and strong stylization strength comparable to that of PhotoWCT. See more results in Supplementary Material.}
    \label{fig:stylized_images}
\end{figure*}

\vspace{-1.1em}\paragraph{Metrics.}
Four no-reference \emph{image quality assessment} algorithms are adopted for the quality evaluation: BRISQUE~\cite{mittal2012no} (\textbf{0}, 100), NIQE~\cite{mittal2012making} (\textbf{0}, $\infty$), NIMA-quality~\cite{talebi2018nima} (1, \textbf{10}), and NIMA-aesthetic (1, \textbf{10}), where the range of a metric value is shown in a parenthesis pair with the bold number as the best. 
We use regularized style loss from DPST~\cite{luan2017deep} to evaluate \emph{stylization strength} (description in the Supplementary Materials). The lower the mean style loss $\mathcal{\bar{L}}_{s,m}$ of a method $m$, the stronger the method's stylization strength.


\vspace{-1.1em}\paragraph{Results.}
Table~\ref{tab:model_charateristics}(c) shows the image quality assessment\footnote{We suspect the NIQE metric is inferior, since the scores for WCT$^2$, PhotoWCT, and our model are better than those for pristine images.} and stylization strength results. Note that all reported scores are the mean of values across the 100 stylized images.

For image quality, we observe that our PhotoWCT$^2$ performs similarly to the other top-performing methods: PhotoWCT and WCT$^2$.  Our PhotoWCT$^2$ performs slightly better with respect to two of the metrics (NIQE and NIMA-a) and slightly worse for the other two metrics.  This highlights that our PhotoWCT$^2$ can preserve the quality of the top-performing PhotoWCT and WCT$^2$ methods while using considerably fewer model parameters.

For stylization strength, we observe again that our PhotoWCT$^2$ performs comparably to the top-performing PhotoWCT; i.e., -0.70 for PhotoWCT versus -0.66 and -0.69 for our PhotoWCT$^2$ (E2E) and PhotoWCT$^2$ (BT).  We attribute the stronger stylization strength of PhotoWCT$^2$ (BT) than that of PhotoWCT$^2$ (E2E) to the better feature/image reconstruction ability of PhotoWCT$^2$ (BT) (see Section~\ref{sec:training_strategies}). Moreover, our models considerably outperform the other two baselines. For example, PhotoWCT$^2$ (BT) results in scores that are $1.71$ and $1.49$ better than that of PhotoNAS and WCT$^2$ respectively. We illustrate qualitatively the stylization strength of our approach compared to the baselines in Figure~\ref{fig:stylized_images}a.

\subsection{Ablation study on training strategies}
\label{sec:training_strategies}
Next we compare using blockwise training to train our AEC$_{bt}$ architecture (described in Section~\ref{sec:training_setting}) with two other methods: end-to-end and vanilla training methods. Recall that end-to-end training enables the decoder to learn the four function inversions at once.  \emph{Vanilla training}, used for WCT$^2$ and PhotoNAS, uses AEC$_{bt}$ as a vanilla autoencoder that only realizes image reconstruction (i.e., without $\textit{reluN\_1}$ feature reproduction) in the decoder.

\vspace{-1.1em}\paragraph{Dataset.}
We randomly sample 500 images from the MSCOCO~\cite{lin2014microsoft} dataset. 

\vspace{-1.1em}\paragraph{Metrics.}
We evaluate with respect to two metrics.  First, we assess the \emph{image reconstruction} ability by computing a pixelwise $L2$ loss between an input image $I$ of shape $H$$\times$$W$$\times$$C$ and a reconstructed image $I_{rec}$: $\frac{||I-I_{rec}||^2_2}{HWC}$.  Second, we assess the \emph{feature reconstruction} ability by computing a relative $L2$ loss between a feature $\mathbf{F}_N$ ($N$ = 1, 2, 3) from the $\textit{reluN\_1}$ layer in the encoder block $\textit{enc}_4\textit{blk}_N$ and a reproduced feature $\mathbf{F}_{N,r}$ from the decoder block $\textit{dec}_{bt}\textit{blk}_{N+1}$: $\frac{||\mathbf{F}_N-\mathbf{F}_{N,r}||^2_2}{||\mathbf{F}_N||^2_2}$. The reason for using the relative error instead of the absolute error is that an element value in $\mathbf{F}_N$ or $\mathbf{F}_{N,r}$ can be large, resulting in a misleading large absolute difference $||\mathbf{F}_N-\mathbf{F}_{N,r}||^2_2$ even if the relative error is small and so suggests a good reconstruction.  Mean values are reported for each training strategy.

\begin{table}[!t]
    \centering
    \small
    \captionsetup{position=top}
    \subfloat[Different training strategies w/ high-freq residuals\vspace{-2mm}]{
    \begin{tabular}{l c c c c}
    \toprule
        Strategy & $\textit{relu3}\_1$ & $\textit{relu2}\_1$ & $\textit{relu1}\_1$ & image \\\cmidrule[0.5pt](r){1-1} \cmidrule[0.5pt](l){2-5}
     Vanilla training & 3.887 & 1.163 & 1.211 & \textbf{0.0003} \\
     End-to-end training & 0.048 & 0.022 & \textbf{0.008} & 0.0008\\
     Blockwise training & \textbf{0.035} & \textbf{0.021} & \textbf{0.008} & 0.0006\\\bottomrule
    \end{tabular}}\\\vspace{2mm}
    \subfloat[Different skip connection types  w/ blockwise training\vspace{-2mm}]{
    \begin{tabular}{l c c c c}
    \toprule
        Type & $\textit{relu3}\_1$ & $\textit{relu2}\_1$ & $\textit{relu1}\_1$ & image \\\cmidrule[0.5pt](r){1-1} \cmidrule[0.5pt](l){2-5}
     no skip connect. & 0.144 & 0.183 & 0.162 & 0.0052\\
     indices of max.\cite{li2018closed} & 0.090 & 0.092 & 0.065 & 0.0028\\
     instance norm.\cite{an2020ultrafast} & 0.045 & 0.048 & 0.037 & 0.0012\\
     wavelet-based skip \cite{yoo2019photorealistic} & 0.043 & 0.030 & 0.010 & \textbf{0.0006}\\
     high-freq residuals & \textbf{0.035} & \textbf{0.021} & \textbf{0.008} & \textbf{0.0006}\\\bottomrule
    \end{tabular}}\vspace{1mm}
    \caption{Loss values for feature and image reconstruction in the decoder resulting from (a) different training strategies and (b) different skip connection types. Results show that blockwise training and high-frequency residuals together achieve the best reconstruction performance.}
    \vspace{-3mm}
    \label{tab:train_w_SCs}
\end{table}


\begin{figure*}[!ht]
    \centering
    \includegraphics[width=0.9\linewidth]{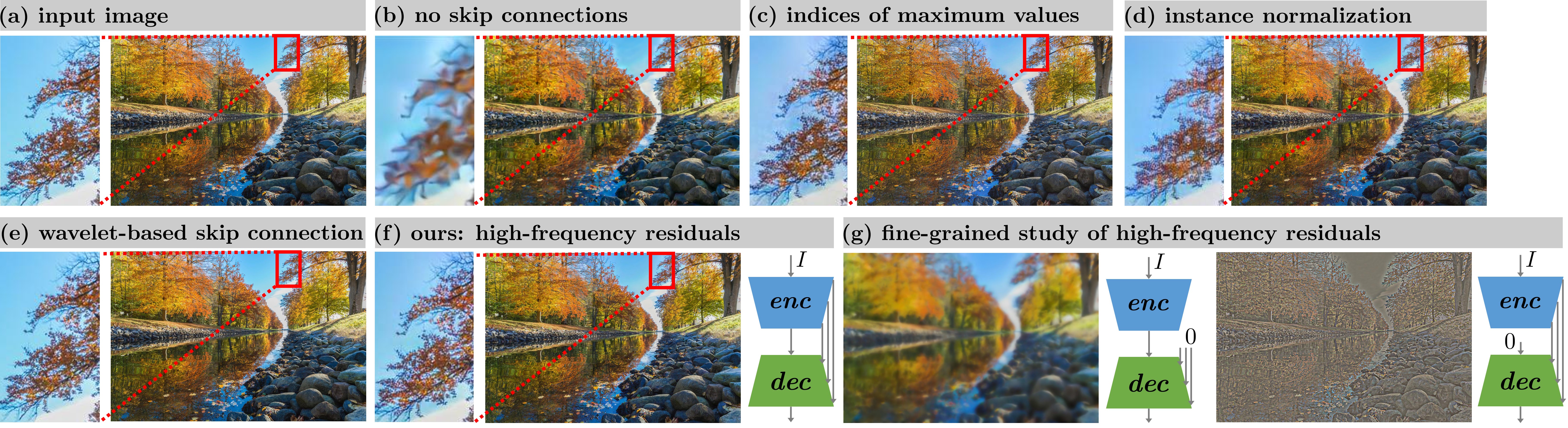}
    \vspace{-3mm}
    \caption{Image reconstruction ability of blockwisely trained autoencoders with different skip connection types. The wavelet-based skip connections in (e) and our high-frequency residuals in (f) achieve the best result among (b-f). The fine grained study in (g) shows its good reconstruction results from the supply of high-frequency information in the input image by the high-frequency residuals. Our high-frequency residuals are the lightweight version of wavelet-based skip connections.}
    \vspace{-3mm}
    \label{fig:reconstruction}
\end{figure*}

\vspace{-1.1em}\paragraph{Results.}
Quantitative results are reported in Table~\ref{tab:train_w_SCs}a.  We observe that our blockwise training results in the best feature reconstruction and second best image reconstruction. We attribute the better reconstruction ability of blockwise training than that of end-to-end training to the distribution of the load of four function inversions into four individual steps. Also, the high feature reconstruction losses of vanilla training indicates that it fails to reproduce any $\textit{reluN\_1}$ features and so fails to realize the PhotoWCT algorithm. Qualitative results from vanilla training are shown in Figure~\ref{fig:stylized_images}b.

\subsection{Ablation study on skip connection types}
\label{sec:bt_with_skip}
Next we compare our skip connections based on high-frequency residuals to three alternative lightweight options (1-3) and the heavyweight option (4): (1) no skip connections, (2) skip connections of indices of maximum values used in PhotoWCT~\cite{li2018closed}, (3) instance-normalized skip connections used in PhotoNAS~\cite{an2020ultrafast}, and (4) wavelet-based skip connections used in WCT$^2$~\cite{yoo2019photorealistic}. To do so, we train four variants of AEC$_{bt}$ with our blockwise training, each with high-frequency residuals replaced by one of the above skip connection types. 
We employ the same experimental metrics and dataset as in Section~\ref{sec:training_strategies}.

\vspace{-1.1em}\paragraph{Results.}
The average loss across all images for feature and image reconstruction is shown in Table~\ref{tab:train_w_SCs}b.  We observe that our skip connections of high-frequency residuals consistently leads to considerable improvements over the three alternative lightweight options. Our skip connections even perform slightly better than the heavyweight wavelet-based skip connections. The autoencoder without skip connections has the worst performance, as indicated by the highest loss values.  Compared to this no skip connection case, our high-frequency residuals improves by $75.7\%$, $88.5\%$, $95.0\%$, and $88.5\%$ for $\textit{relu3\_1}$, $\textit{relu2\_1}$, $\textit{relu1\_1}$, and image reconstruction, respectively.  When skip connections of indices of maximal values or instance-normalized skip connections are applied, the performance improves slightly.


In Figure~\ref{fig:reconstruction}, we exemplify the image reconstruction ability of each model with different skip connection types. We notice in Figure~\ref{fig:reconstruction}b that the reconstructed result without skip connections captures the general idea, such as color, of the input image but fails to reconstruct the high-frequency edges. Although skip connections of indices of maximum values and instance-normalized skip connections improve the reconstruction of high-frequency components, failures still arise including uneven blue sky around leaves and artifacts within leaves as shown in the zoomed-in crops in Figures~\ref{fig:reconstruction}c and \ref{fig:reconstruction}d. In contrast, the results from wavelet-based skip connections and high-frequency residuals in Figures~\ref{fig:reconstruction}(ef) are reconstructed better and closer to the original image.

We exemplify the influence of high-frequency residuals on image reconstruction in Figure~\ref{fig:reconstruction}g. We show what a reconstructed image looks like when there are no high-frequency residuals and when there are only high-frequency residuals. In the former case, the skip connections to the decoder are replaced by zero tensors, while in the latter case, the connection between the encoder and decoder at the bottleneck is cut and instead the input to the decoder is a zero tensor. As shown in Figure~\ref{fig:reconstruction}g, the result from no high-frequency residuals is a low-frequency, blurry image, while the result from high-frequency residuals is only the high-frequency edges of the input image. This implies that high-frequency residuals reinforce image reconstruction by supplying high-frequency components from the input image.

\section{Conclusion}
\vspace{-2mm}
We designed a compact photorealistic style transfer model to embed lightweight representations of PhotoWCT's coarse-to-fine feature transformations and WCT$^2$'s wavelet-based skip connections.  Two novel mechanisms, blockwise training and skip connections of high frequency residuals made this compact representation possible.  Experiments demonstrate that our PhotoWCT$^2$ preserves the strong stylization strength of PhotoWCT and good image/feature reconstruction ability of WCT$^2$ while supporting stylization of higher resolution images and faster stylization speed without loss to image quality.

\section*{Supplementary Materials}
This document supplements Sections 3.1, 3.2, and 4.2 of the main paper. It includes the following:

\begin{itemize}[leftmargin=*]
\item Details of ZCA transformation (supplements \textbf{Section 3.1}).
\item Description of the bug in the post-processing code of PhotoWCT, our modification, and why fixing the bug greatly improves the speed of the post-processing used in PhotoWCT (supplements \textbf{Sections 3.1}).
\item Additional details about our blockwise model architecture and our training strategy (supplements \textbf{Section 3.2}).
\item Results of blockwise training in the reversed order (supplements \textbf{Section 3.2.2}).
\item PhotoWCT with three cascaded autoencoders and its redesign (supplements \textbf{Section 3.2.2}).
\item Additional details about our regularized style loss (supplements \textbf{Section 4.2}).
\item Qualitative stylization results (supplements \textbf{Section 4.2}).
\end{itemize}

\subsection*{Details of ZCA transformation}
ZCA transformation~\cite{li2017universal} is the key to the realization of style transfer in WCT$^2$, PhotoWCT, PhotoNAS, and our PhotoWCT$^2$ by making the gram matrix of a content feature match that of a style feature. It takes as input a content feature of shape $H_c\times W_c\times C$ and a style feature of shape $H_s\times W_s\times C$ extracted from, say, the $\textit{reluN\_1}$ layer of VGGNet, where $H_c$ ($H_s$) and $W_c$ ($W_s$) are the height and width of the content (style) feature, while $C$ is the channel length. We first reshape the content and style features to the shapes $C\times H_c W_c$ and $C\times H_s W_s$ and denote the reshaped features $\mathbf{F}_c$ and $\mathbf{F}_s$, respectively. Then we apply eigen-decomposition to the covariances $\frac{1}{H_c W_c}\mathbf{\bar{F}}_c\mathbf{\bar{F}}_c^\mathrm{T}$ and $\frac{1}{H_s W_s}\mathbf{\bar{F}}_s\mathbf{\bar{F}}_s^\mathrm{T}$:
\begin{equation}
    \begin{aligned}
        &\frac{1}{H_c W_c}\mathbf{\bar{F}}_c\mathbf{\bar{F}}_c^\mathrm{T} = \mathbf{E}_c\mathbf{\Lambda}_c\mathbf{E}_c^\mathrm{T} \\
        &\frac{1}{H_s W_s}\mathbf{\bar{F}}_s\mathbf{\bar{F}}_s^\mathrm{T} =
        \mathbf{E}_s\mathbf{\Lambda}_s\mathbf{E}_s^\mathrm{T},
    \end{aligned}
\end{equation}
where $\mathbf{\bar{F}}_c$ and $\mathbf{\bar{F}}_s$ are the centralized content and style features: 
\begin{equation}
    \begin{aligned}
        &\mathbf{\bar{F}}_c = \mathbf{F}_c - \mathrm{mean}(\mathbf{F}_c) =\mathbf{F}_c - \sum_{i=1}^{H_c W_c} [\mathbf{F}_c]_{:,i} \\
        &\mathbf{\bar{F}}_s = \mathbf{F}_s - \mathrm{mean}(\mathbf{F}_s) =\mathbf{F}_s - \sum_{i=1}^{H_s W_s} [\mathbf{F}_s]_{:,i}.
    \end{aligned}
\end{equation}
By transforming the content feature $\mathbf{F}_c$ as in Equation~\ref{eq:zca}, the gram matrix of the transformed feature $\mathbf{F}_{cs}$ will match that of the style feature. 
\begin{equation}
    \mathbf{F}_{cs} = (\mathbf{E}_s\mathbf{\Lambda}^{\frac{1}{2}}_s\mathbf{E}_s^\mathrm{T})(\mathbf{E}_c\mathbf{\Lambda}^{-\frac{1}{2}}_c\mathbf{E}_c^\mathrm{T})\mathbf{\bar{F}}_c + \mathrm{mean}(\mathbf{F}_s)
    \label{eq:zca}
\end{equation}
It can be shown that $\frac{1}{H_c W_c}\mathbf{F}_{cs}\mathbf{F}_{cs}^\mathrm{T}$ is equal to $\frac{1}{H_s W_s}\mathbf{F}_s\mathbf{F}_s^\mathrm{T}$~\cite{chiu2019understanding}.

\begin{figure*}[!t]
    \centering
    \includegraphics[width=\textwidth]{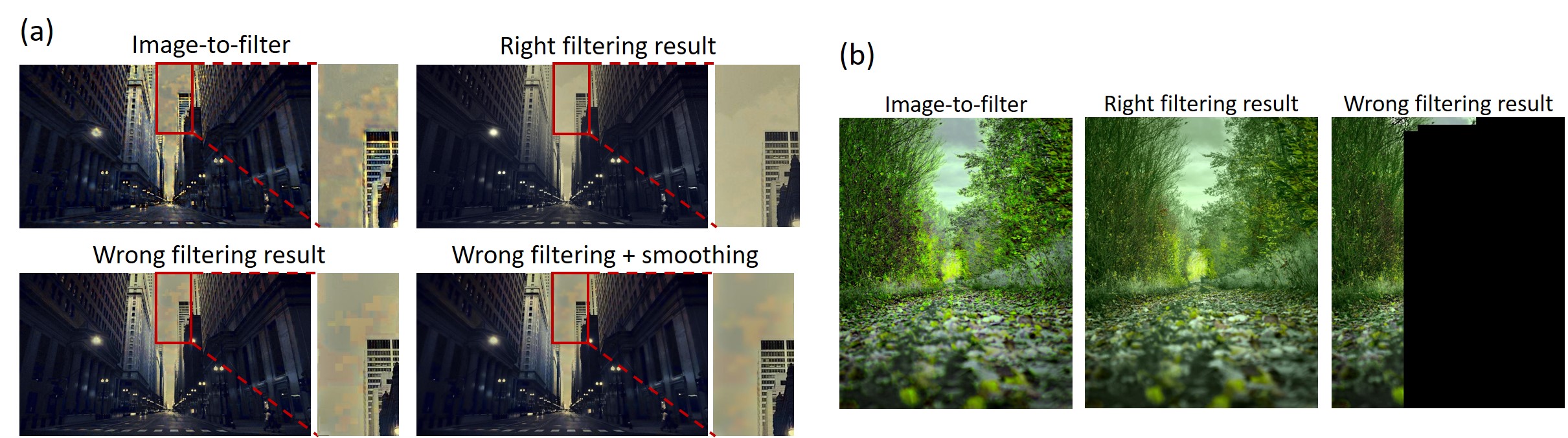}
    \caption{Two examples showing the improvement after fixing the bug in the PhotoWCT code. In (a), we observe that the result from the wrong invocation of the \texttt{guidedFilter} function results in a grid artifact that needs to be fixed with additional smoothing. (b) shows a worse case where wrongly invoking the function results in a failure. }
    \label{fig:guided_filter}
\end{figure*}

\begin{figure*}[t!]
    \centering
    \includegraphics[width=0.95\textwidth]{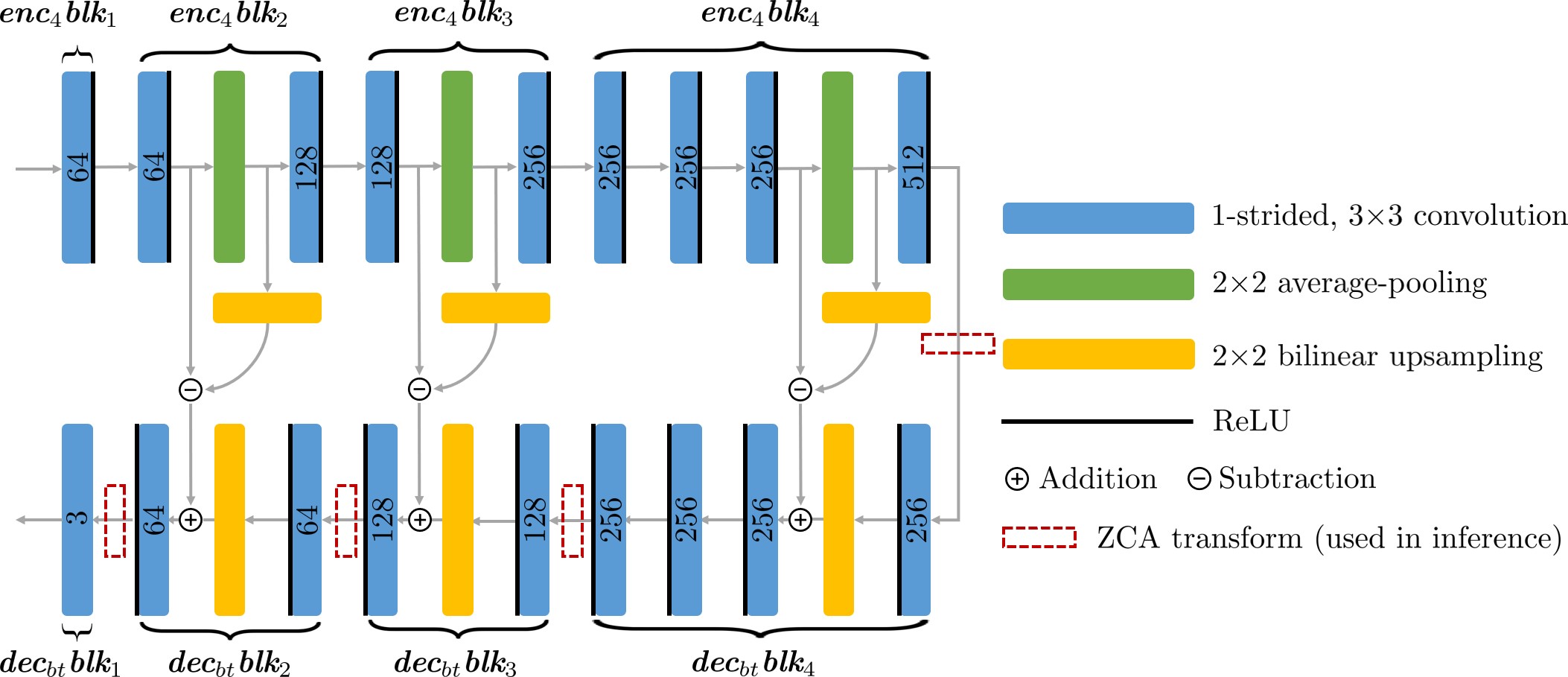}
    \caption{Detailed structure of our model's architecture that gets used during blockwise training. The number in each convolution layer is the number of channels at its output.}
    \label{fig:model}
\end{figure*}

\vspace{1em}\subsection*{Modification of PhotoWCT's post-processing code}
The post-processing in the PhotoWCT code includes two parts: guided filtering and image smoothing, with the former code running much faster than the latter. Since smoothing is also one effect of guided filtering, the second post-processing step of image smoothing is not necessary. We suspect the image smoothing function was invoked to overcome a misuse of the $\texttt{guidedFilter}$ function provided in the OpenCV package, as described below. 

The $\texttt{guidedFilter}$ function takes four arguments: a guide image, an image to filter, a filtering window radius, and an $\epsilon$ parameter to prevent overfitting. The PhotoWCT code follows the guided filter paper~\cite{he2010guided} to use $0.1^2$ as the value of $\epsilon$\footnote{In practice, values between $0.01^2$ to $0.1^2$ work well.}. However, while the paper assumes image pixel values are in the range of $0$ and $1$, $\texttt{guidedFilter}$ takes images with pixel values ranged from $0$ to $255$ and so the value of $\epsilon$ should be scaled accordingly. In our modification, we set $\epsilon$ to be $(0.02\times 255)^2$.  As expected, with this value, the second post-processing step of image smoothing becomes unnecessary, and the removal of the second post-processing greatly improves the speed reported in previous works.

Figure~\ref{fig:guided_filter} exemplifies the importance of fixing the code. Before fixing the bug, a filtering result might suffer from a grid artifact, where a region that is supposed to be smooth contains groups of pixels that are not well blended.  This artifact is exemplified in Figure~\ref{fig:guided_filter}(a). To remove this artifact, the second post-processing step of smoothing is needed. Even still, this extra filtering step can fail to conceal the issues introduced by the bug, as exemplified in Figure~\ref{fig:guided_filter}(b).

We set the filtering window radius to be $100$ for the speed test in Section~4.1.  We set it to $50$ for evaluation of the image quality and stylization strength in Section~4.2, since the test images used in the speed test are (much) larger than the images from the DPST dataset used in Section~4.2.

\vspace{1em}\subsection*{Model-to-train and training details}
A detailed diagram showing the structure of our model that gets used during blockwise training is shown in Figure~\ref{fig:model}, expanding upon its illustration in Figure 1b:4 and Figure 3a of the main paper. For training, we use the MS-COCO dataset~\cite{lin2014microsoft}. Each image in the dataset is resized to 512$\times$512 and randomly cropped to 256$\times$256.  We use a batch size of eight images. We use the Adam optimizer with learning rate $1 \times 10^{-4}$ and without weight decay. In blockwise training, each $\textit{dec}_{bt}\textit{blk}_N$ block is trained for 20 epoch.

\begin{figure}
    \centering
    \includegraphics[height=0.25\textheight]{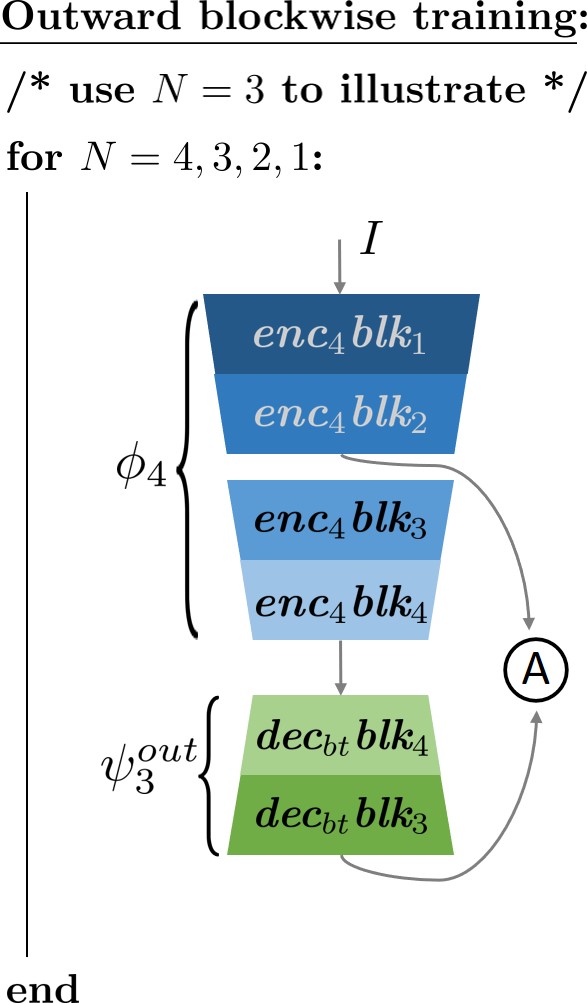}
    \caption{Outward blockwise training}
    \label{fig:outward_bt}
\end{figure}

\setlength{\tabcolsep}{3pt}
\begin{table*}[t!]
    \centering\small
    \begin{tabular}{l c c c c c c c c c c c c }
    \toprule
        \multirow{3}{*}{Model} & \multicolumn{2}{c}{(a) Size} & \multicolumn{5}{c}{(b) Speed performance} & \multicolumn{5}{c}{(c) Image quality \& Stylization strength} \\
    \cmidrule[0.5pt](lr){2-3}\cmidrule[0.5pt](lr){4-8}\cmidrule[0.5pt](lr){9-13}
         & \multirow{2}{*}{\# par} & \multirow{2}{*}{\# layer} & \multirow{2}{*}{1024$\times$512} & HD & FHD & QHD & 4K & BRIS & NIQE & NIMA-q & NIMA-a & \multirow{2}{*}{$\bar{\mathcal{L}}_{s,m}$} \\
         & & & &1280$\times$720 & 1920$\times$1080 & 2560$\times$1440 & 3840$\times$2160 & (27.4) & (3.19) & (5.11) & (5.27) & \\
    \midrule
    PhNAS & 40.24M & 35 & 0.23 & OOM & OOM & OOM & OOM & 33.0 & 3.24 & 4.75 & 4.92 & 1.42\\
    WCT$^2$ & 10.12M & \bf{24} & 0.30 & 0.43 & 0.80 & OOM & OOM & \bf{30.8} & 3.07 & \bf{4.91} & 5.01 & 1.15\\
    PhWCT & 8.35M & 48 & 0.21+0.03 & 0.32+0.06 & 0.61+0.14 & 1.01+0.23 & OOM & 31.8 & \bf{2.90} & 4.88 & 5.06 & \bf{-0.57} \\
    AEC$_{e2e}$ & \bf{7.05M} & \bf{24} & \bf{0.18+0.03}  & \bf{0.24+0.06} & \bf{0.39+0.14} & \bf{0.59+0.23} & \bf{1.22+0.54} & 31.7 & 2.91 & 4.90 & \bf{5.10} & -0.52\\
    AEC$_{bt}^{in}$ & \bf{7.05M} & \bf{24} & \bf{0.18+0.03}  & \bf{0.24+0.06} & \bf{0.39+0.14} & \bf{0.59+0.23} & \bf{1.22+0.54} & 31.6 & \bf{2.90} & 4.90 & \bf{5.10} & -0.55 \\
    AEC$_{bt}^{out}$ & \bf{7.05M} & \bf{24} & \bf{0.18+0.03}  & \bf{0.24+0.06} & \bf{0.39+0.14} & \bf{0.59+0.23} & \bf{1.22+0.54} & 31.8 & 2.92 & 4.88 & 5.06 & -0.54\\
    \midrule\midrule
    PhWCT3 & 1.34M & 24 & 0.13+0.03 & 0.19+0.06 & 0.36+0.14 & 0.60+0.23 & OOM & 32.1 & 2.98 & \bf{4.88} & 5.04 & \bf{-0.47} \\
    AEC$_{bt,3}^{in}$ & \bf{1.15M} & \bf{14} & \bf{0.09+0.03}  & \bf{0.13+0.06} & \bf{0.24+0.14} & \bf{0.40+0.23} & \bf{0.77+0.54} & \bf{31.6} & \bf{2.92} & \bf{4.88} & \bf{5.06} & -0.42\\
    \bottomrule
    \end{tabular}
    \vspace{-2mm}
    \caption{Extended Table 1 from the main paper. (Top 6 rows) Performance comparison between three previous methods PhotoNAS, WCT$^2$, and PhotoWCT and our models AEC$_{e2e}$, AEC$_{bt}^{in}$ and AEC$_{bt}^{out}$ (PhotoWCT$^2$ resulting from end-to-end training, inward and outward blockwise trainings). (Last 2 rows) Comparison of PhotoWCT3 (PhotoWCT reduced to three cascaded autoencoders) and AEC$_{bt,3}^{in}$ (redesign of PhotoWCT3 using inward blockwise training).}
    \label{tab:extended_table1}
\end{table*}

\vspace{1em}\subsection*{Results of blockwise training in the reversed order}
The blockwise training in the paper trains the decoder blocks $\textit{dec}_{bt}\textit{blk}_N$'s in the order from $N=1$ to $N=4$. That is, the training is inward from the outermost block $\textit{dec}_{bt}\textit{blk}_1$ to the innermost $\textit{dec}_{bt}\textit{blk}_4$. Here we show the results of outward training from the innermost to the outermost block as exemplified in Figure~\ref{fig:outward_bt}. 

In outward blockwise training, the decoder blocks $\textit{dec}_{bt}\textit{blk}_N$'s are trained in the order from $N=4$ to $N=1$ by minimizing the function inversion loss $\mathcal{L}_N^{out}$:
\begin{equation}
    \mathcal{L}_N^{out}(I) =
      ||\phi_{N-1}(I)-\psi^{out}_N(\phi_4(I))||^2_2
    \label{eq:loss_outward}
\end{equation}
where $\phi_N$ and $\psi^{out}_N$ are the functions of the series $\{\textit{enc}_4\textit{blk}_1, \dots, \textit{enc}_4\textit{blk}_N\}$ and $\{\textit{dec}_{bt}\textit{blk}_4, \dots, \textit{dec}_{bt}\textit{blk}_N\}$, respectively. When training a decoder block, the previously trained blocks and the encoder are fixed. Note that when $N=1$, $\mathcal{L}_1^{out}$ minimizes the reconstruction loss.  In this document, we denote our models trained outwardly and inwardly as AEC$_{bt}^{out}$ and AEC$_{bt}^{in}$ (PhotoWCT$^2$ (BT)), respectively.


\begin{table}[!t]
    \centering
    \small
    \captionsetup{position=top}
    \begin{tabular}{l c c c c}
    \toprule
        Strategy & $\textit{relu3}\_1$ & $\textit{relu2}\_1$ & $\textit{relu1}\_1$ & image \\\cmidrule[0.5pt](r){1-1} \cmidrule[0.5pt](l){2-5}
     Outward & 0.037 & 0.028 & 0.009 & 0.0010\\
     Inward & \textbf{0.035} & \textbf{0.021} & \textbf{0.008} & \bf{0.0006}\\\bottomrule
    \end{tabular}\\\vspace{1mm}
    \caption{Loss values for feature and image reconstruction in the decoders resulting from inward and outward blockwise trainings. The inward training has a slightly better image and feature reconstruction ability than the outward training.}
    \vspace{-4mm}
    \label{tab:extended_table2}
\end{table}

Next we compare the inward and the outward blockwise trainings using the same metrics in the main paper. Table~\ref{tab:extended_table1}(c) shows the mean metric scores for image quality and stylization strength across 100 stylized images resulting from each method. Note that the values of $\mathcal{\bar{L}}_{s,m}$ reported in Table~\ref{tab:extended_table1} here are different from those reported in the Table 1 in the main paper, since here style losses of AEC$_{bt}^{out}$, PhotoWCT3, and AEC$_{bt,3}^{in}$ (see next section) are included in the normalization for the computation of $\mathcal{\bar{L}}_{s,m}$. We observe both AEC$_{bt}^{in}$ (i.e., PhotoWCT$^2$ (BT)) and AEC$_{bt}^{out}$ result in image quality comparable to that from the baselines and as strong stylization strength as PhotoWCT. Table~\ref{tab:extended_table2} shows the loss values averaged across 500 images for feature and image reconstruction in the decoders trained by inward and outward blockwise trainings. We observe the inward training has a slightly better image and feature reconstruction ability than the outward training. 

\vspace{1em}\subsection*{Redesign of PhotoWCT to use three autoencoders}
As discussed in the main paper, our method can generalize for use with different numbers of autoencoders.  We illustrate this here by removing the fourth autoencoder AEC$_4$ in the PhotoWCT cascade. We denote the resulting 3-autoencoder PhotoWCT as PhotoWCT3. 

Quantitative results comparing PhotoWCT and PhotoWCT3 are shown in Table~\ref{tab:extended_table1}. As expected, we observe a slight drop in stylization strength from PhotoWCT (-0.57) to PhotoWCT3 (-0.47), while PhotoWCT3 uses fewer parameters and runs at faster speeds.

We next apply our inward blockwise training method to redesign PhotoWCT3 into a single autoencoder, which is denoted AEC$_{bt,3}^{in}$ here. Results are shown in Table~\ref{tab:extended_table1}.  Reinforcing our findings in the main paper, we observe that AEC$_{bt,3}^{in}$ achieves comparable stylization strength while yielding faster speeds than PhotoWCT3. For example, it takes 0.2 fewer seconds for QHD rendering for AEC$_{bt,3}^{in}$ than PhotoWCT3. When comparing AEC$_{bt,3}^{in}$ to PhotoWCT, while there is a more noticeable drop in stylization strength from the original PhotoWCT method (i.e., four autoencoders) to AEC$_{bt,3}^{in}$, AEC$_{bt,3}^{in}$ runs much faster than PhotoWCT. Take QHD rendering for instance. AEC$_{bt,3}^{in}$ spends only half the rendering time of PhotoWCT (0.64s vs. 1.24s). 

Note that when redesigning PhotoWCT3 into a single autoencoder, we use inward rather than outward blockwise training. The reason is that inward blockwise training not only results in better image and feature reconstruction ability, but while we redesign PhotoWCT using inward blockwise training as in the main paper, we already accomplish the redesign of PhotoWCT3. That is, inward blockwise training trains the decoder blocks $\textit{dec}_{bt}\textit{blk}_1$, $\textit{dec}_{bt}\textit{blk}_2$, and $\textit{dec}_{bt}\textit{blk}_3$ in AEC$_{bt,3}^{in}$ first and then trains the final block $\textit{dec}_{bt}\textit{blk}_4$ used in AEC$_{bt}^{in}$. In contrast, due to the reversed training order, redesigning PhotoWCT3 with outward blockwise training requires re-training from scratch.

\vspace{1em}\subsection*{Regularized style loss}
We use the regularized style loss $\mathcal{L}_{s}$ in equation~\ref{eq:reg_style_loss}, which was introduced in DPST~\cite{luan2017deep}, for the evaluation of stylization strength. Intuitively, style can be thought of as a composition of ingredients such as color, lightness, and artistic effects, including image pattern and painting styles (oil paintings, watercolor paintings, etc.). These ingredients are captured in Gatys et al.'s~\cite{gatys2016image} formulation of style loss ($\sum \beta_l \mathcal{L}_{s,l}$). However, since artistic effects result in artifacts, they are undesired ingredients in photorealistic stylization. To avoid these artifacts, a regularization term $\mathcal{L}_{reg}$ is introduced to Gatys's formulation to remove the artistic effects. Mathematically, with $I_o$, $I_c$, $I_s$ being the stylized, content, and style images, the regularized loss is defined as follows:
\begin{equation}
    \mathcal{L}_{s}(I_o; I_c, I_s) = \sum_{l=1}^5 \beta_l \mathcal{L}_{s,l}(I_o, I_s) + \lambda \mathcal{L}_{reg}(I_o, I_c), 
    \label{eq:reg_style_loss}
\end{equation}
\begin{equation}
    \mathcal{L}_{s,l}(I_o, I_s) = ||\frac{1}{H_{o,l}W_{o,l}}\mathbf{F}_{o,l}\mathbf{F}_{o,l}^\mathrm{T} - \frac{1}{H_{s,l}W_{s,l}}\mathbf{F}_{s,l}\mathbf{F}_{s,l}^\mathrm{T}||_2^2,
    \label{eq:style_loss}
\end{equation}
\begin{equation}
    \mathcal{L}_{reg}(I_o, I_c) = \sum_{ch\in \{R,G,B\}} vec(I_{o,ch})^\mathrm{T}\mathcal{M}(I_{c,ch})vec(I_{o,ch}). 
    \label{eq:reg_loss}
\end{equation}
In equation~\ref{eq:style_loss}, $\mathbf{F}_{o,l}$ and $\mathbf{F}_{s,l}$ are the $\textit{relu'l'\_1}$ features of $I_o$ and $I_s$ extracted from VGGNet. $(H_{o,l}, W_{o,l})$ and $(H_{s,l}, W_{s,l})$ are the $($height, width$)$ of $\mathbf{F}_{o,l}$ and $\mathbf{F}_{s,l}$. In equation~\ref{eq:reg_loss}, $vec(I_{o,ch})$ is the pixels of $I_o$ in the $ch$ channel vectorized into a column vector. $\mathcal{M}(I_{c,ch})$ is the Matting Laplacian matrix of the $ch$ channel of $I_c$. Following the official implementation of DPST, the weights $\beta_l$'s are $1/5$, while $\lambda$ is $10^2$.

To account for regularized style loss values falling in different ranges for different content-style pairs, we normalize the loss value $\mathcal{L}_{s,m,p}$ resulting from the photorealistic style transfer method $m$ and the content-style pair $p$ as follows ($m$ $\in$ $\{$$\text{PhotoNAS}$, $\text{WCT}^2$, $\text{PhotoWCT}$, AEC$_{e2e}$, AEC$_{bt}^{in}$$\}$ in the main paper, while $m$ $\in$ $\{$$\text{PhotoNAS}$, $\text{WCT}^2$, $\text{PhotoWCT}$, AEC$_{e2e}$, AEC$_{bt}^{in}$, AEC$_{bt}^{out}$, $\text{PhotoWCT}3$, AEC$_{bt,3}^{in}$$\}$ in this document):
\begin{equation}
   \mathcal{\bar{L}}_{s,m,p} = \frac{\mathcal{L}_{s,m,p}-\mu_p}{\sigma_p} = \frac{\mathcal{L}_{s,m,p} - \frac{1}{|m|}\sum_{m}{\mathcal{L}_{s,m,p}}}{\sqrt{\frac{1}{|m-1|}\sum_{m}(\mathcal{L}_{s,m,p}-\mu_p)^2}},
   \label{eq:norm_style_loss}
\end{equation}
where $|m|$ is the number of considered methods (i.e., $m$ is five and eight in the main paper and this document, respectively.)
As such, $\mathcal{\bar{L}}_{s,m,p}$ is distributed around $0$. Moreover, given a content-style pair $p$, the relative order of $\mathcal{\bar{L}}_{s,m,p}$'s is preserved to match that of $\mathcal{L}_{s,m,p}$'s. Note that the reported loss value $\mathcal{\bar{L}}_{s,m}$ for the method $m$ is the mean across 100 normalized style losses $\{\mathcal{\bar{L}}_{s,m,1}, \dots, \mathcal{\bar{L}}_{s,m,100}\}$ for 100 stylized images resulting from $m$.

\vspace{1em}\subsection*{Stylization results}
\begin{figure*}
    \centering
    \includegraphics[width=1.0\textwidth]{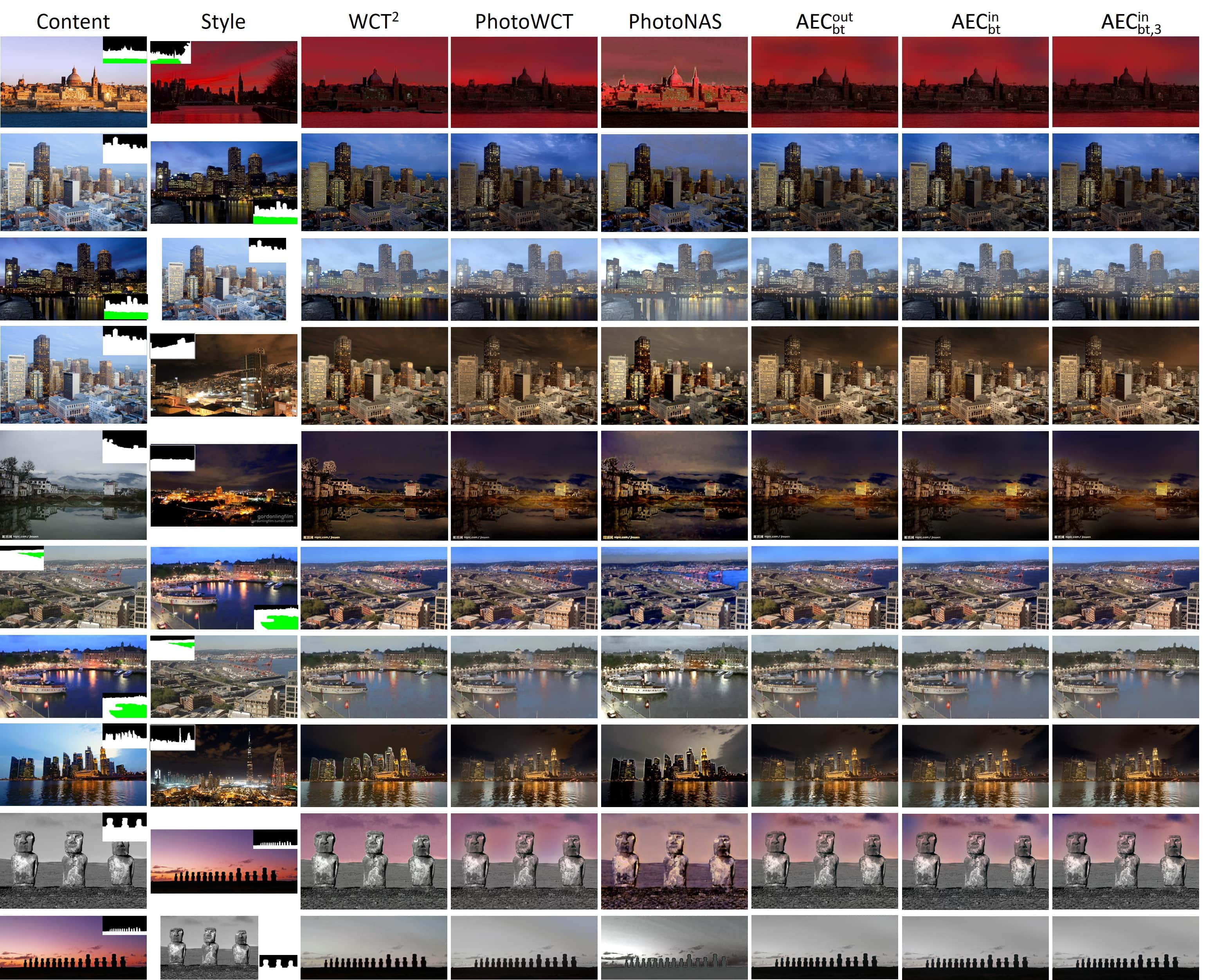}
    \caption{Results of stylization with segmentation for images in the DPST dataset. Three baselines are two state-of-the-arts PhotoWCT~\cite{li2018closed} and WCT$^2$~\cite{yoo2019photorealistic} and a more recent method PhotoNAS~\cite{an2020ultrafast}, while AEC$_{bt}^{in}$ (i.e., PhotoWCT$^2$) and AEC$_{bt}^{out}$ are the autoencoder in Figure~\ref{fig:model} trained inward blockwisely and outward blockwisely, respectively, and AEC$_{bt,3}^{in}$ is AEC$_{bt}^{in}$ with $\text{enc}_4\text{blk}_4$ and $\text{dec}_{bt}\text{blk}_4$ removed. The results exemplify that our models achieve comparable stylization performance to the state-of-the-arts in a fraction of the time required by the existing methods. (Part 1/11) }
    \label{fig:stylization1}
\end{figure*}

\begin{figure*}
    \centering
    \includegraphics[width=1.0\textwidth]{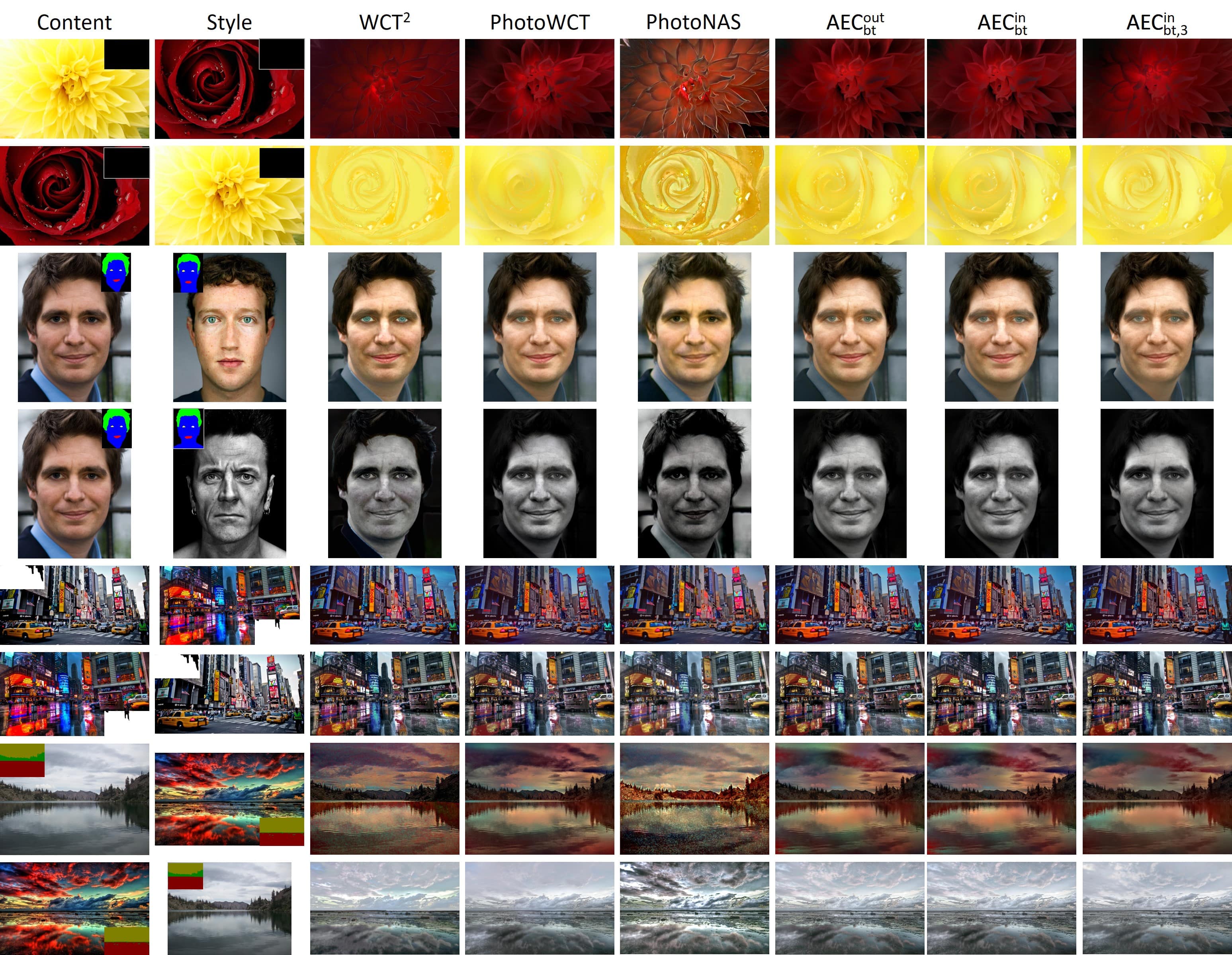}
    \caption{Results of stylization with segmentation for images in the DPST dataset. Three baselines are two state-of-the-arts PhotoWCT~\cite{li2018closed} and WCT$^2$~\cite{yoo2019photorealistic} and a more recent method PhotoNAS~\cite{an2020ultrafast}, while AEC$_{bt}^{in}$ (i.e., PhotoWCT$^2$) and AEC$_{bt}^{out}$ are the autoencoder in Figure~\ref{fig:model} trained inward blockwisely and outward blockwisely, respectively, and AEC$_{bt,3}^{in}$ is AEC$_{bt}^{in}$ with $\text{enc}_4\text{blk}_4$ and $\text{dec}_{bt}\text{blk}_4$ removed. The results exemplify that our models achieve comparable stylization performance to the state-of-the-arts in a fraction of the time required by the existing methods. (Part 2/11) }
    \label{fig:stylization2}
\end{figure*}

\begin{figure*}
    \centering
    \includegraphics[width=1.0\textwidth]{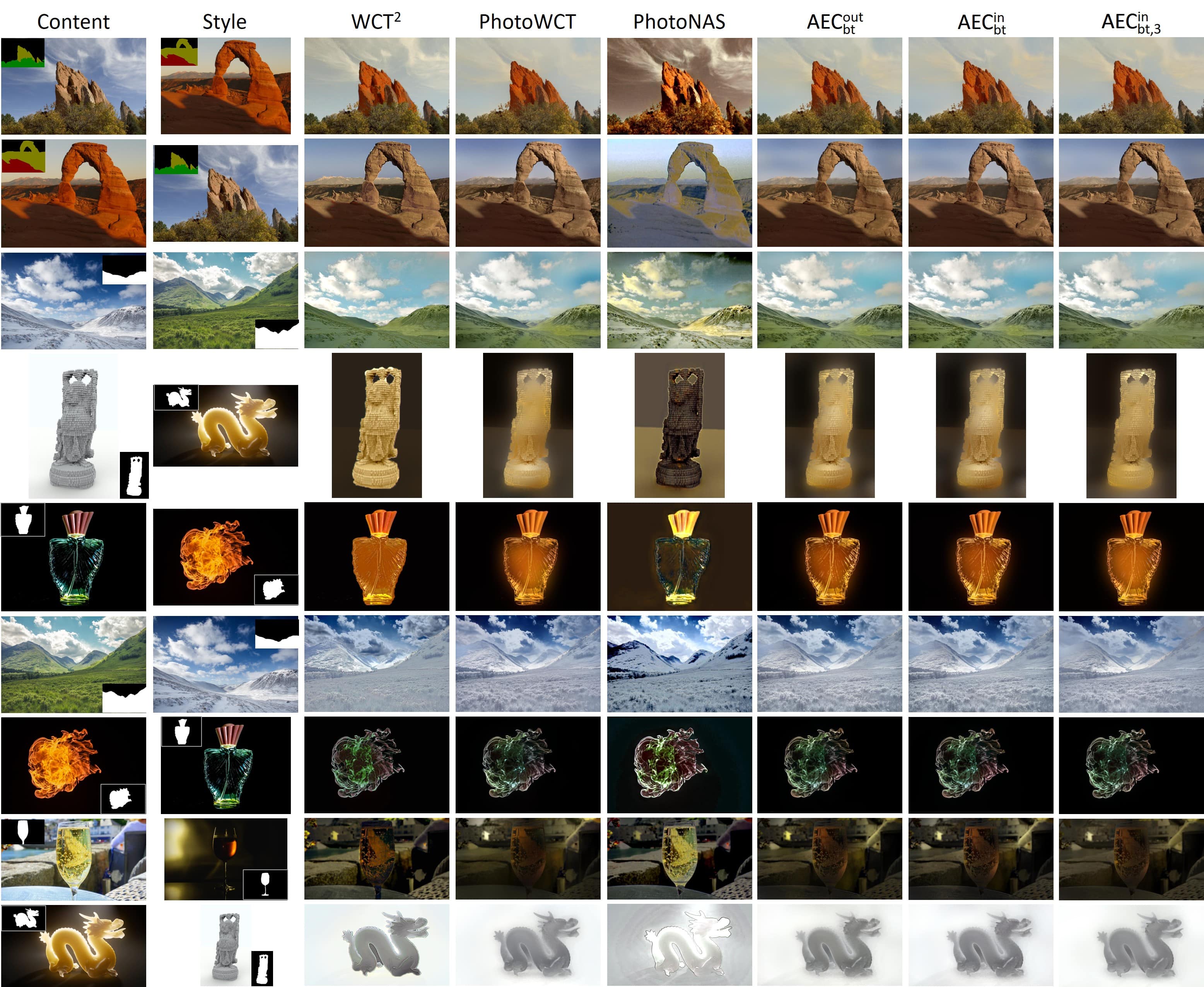}
    \caption{Results of stylization with segmentation for images in the DPST dataset. Three baselines are two state-of-the-arts PhotoWCT~\cite{li2018closed} and WCT$^2$~\cite{yoo2019photorealistic} and a more recent method PhotoNAS~\cite{an2020ultrafast}, while AEC$_{bt}^{in}$ (i.e., PhotoWCT$^2$) and AEC$_{bt}^{out}$ are the autoencoder in Figure~\ref{fig:model} trained inward blockwisely and outward blockwisely, respectively, and AEC$_{bt,3}^{in}$ is AEC$_{bt}^{in}$ with $\text{enc}_4\text{blk}_4$ and $\text{dec}_{bt}\text{blk}_4$ removed. The results exemplify that our models achieve comparable stylization performance to the state-of-the-arts in a fraction of the time required by the existing methods. (Part 3/11) }
    \label{fig:stylization3}
\end{figure*}

\begin{figure*}
    \centering
    \includegraphics[width=1.0\textwidth]{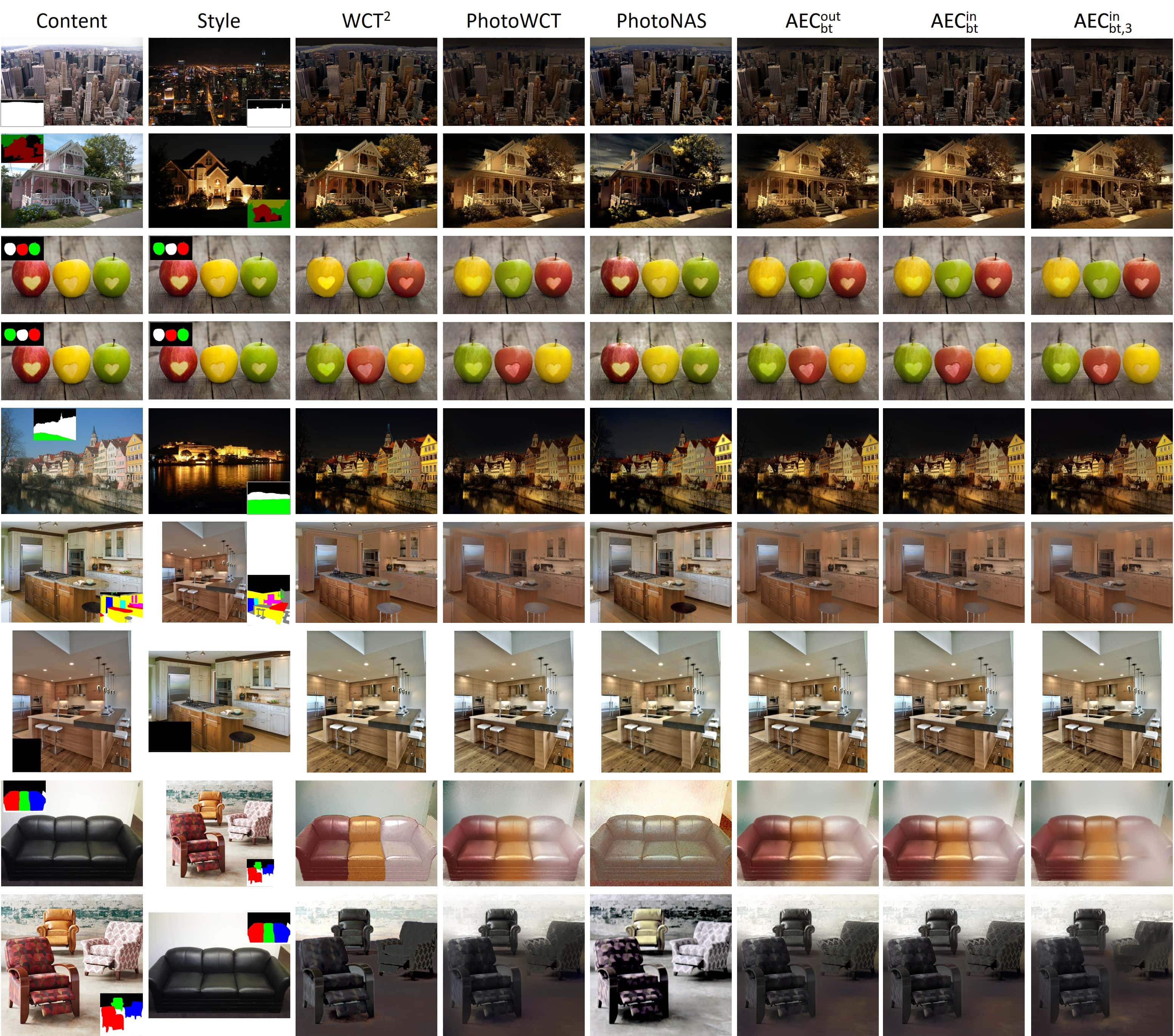}
    \caption{Results of stylization with segmentation for images in the DPST dataset. Three baselines are two state-of-the-arts PhotoWCT~\cite{li2018closed} and WCT$^2$~\cite{yoo2019photorealistic} and a more recent method PhotoNAS~\cite{an2020ultrafast}, while AEC$_{bt}^{in}$ (i.e., PhotoWCT$^2$) and AEC$_{bt}^{out}$ are the autoencoder in Figure~\ref{fig:model} trained inward blockwisely and outward blockwisely, respectively, and AEC$_{bt,3}^{in}$ is AEC$_{bt}^{in}$ with $\text{enc}_4\text{blk}_4$ and $\text{dec}_{bt}\text{blk}_4$ removed. The results exemplify that our models achieve comparable stylization performance to the state-of-the-arts in a fraction of the time required by the existing methods. (Part 4/11) }
    \label{fig:stylization4}
\end{figure*}

\begin{figure*}
    \centering
    \includegraphics[width=1.0\textwidth]{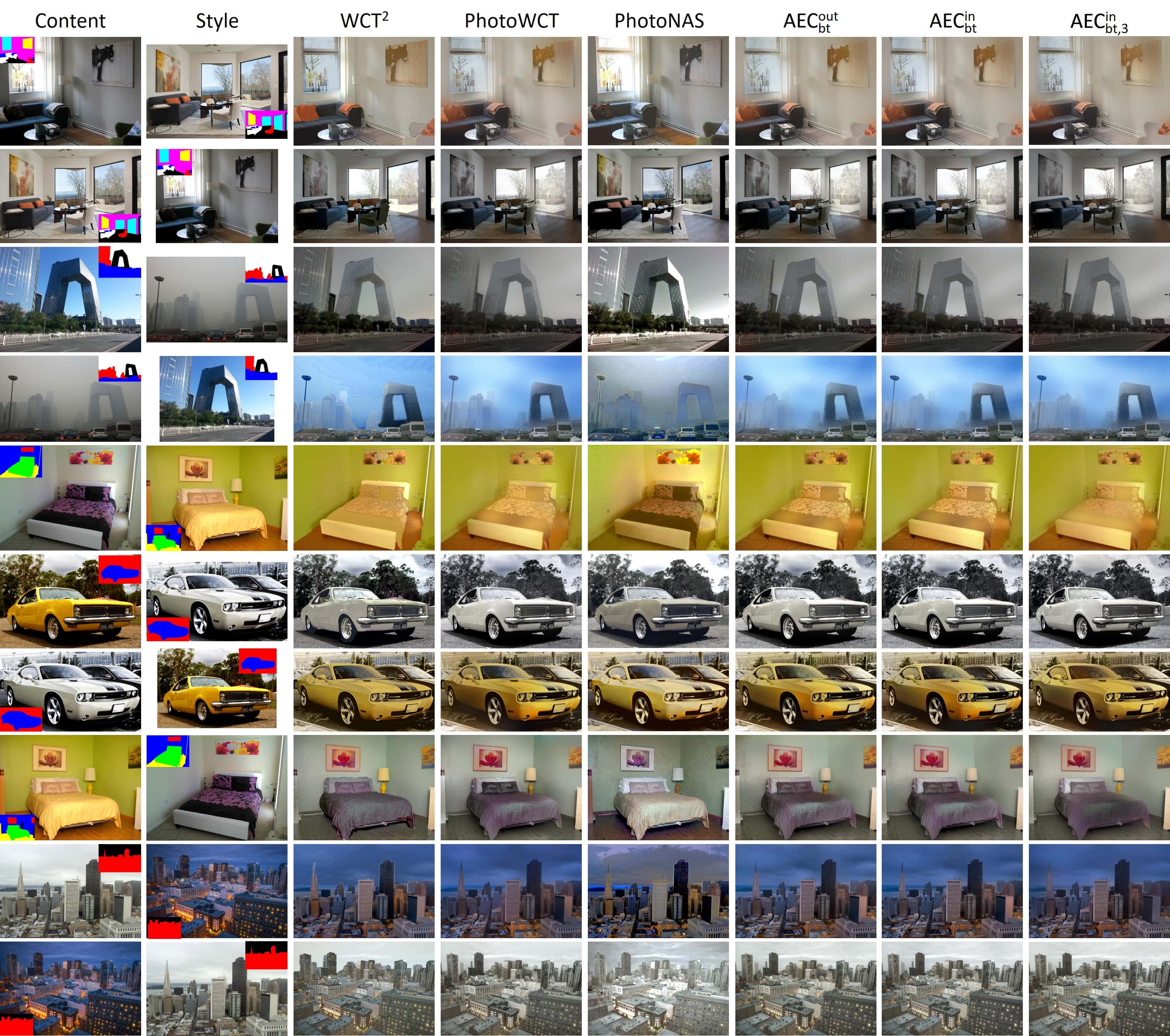}
    \caption{Results of stylization with segmentation for images in the DPST dataset. Three baselines are two state-of-the-arts PhotoWCT~\cite{li2018closed} and WCT$^2$~\cite{yoo2019photorealistic} and a more recent method PhotoNAS~\cite{an2020ultrafast}, while AEC$_{bt}^{in}$ (i.e., PhotoWCT$^2$) and AEC$_{bt}^{out}$ are the autoencoder in Figure~\ref{fig:model} trained inward blockwisely and outward blockwisely, respectively, and AEC$_{bt,3}^{in}$ is AEC$_{bt}^{in}$ with $\text{enc}_4\text{blk}_4$ and $\text{dec}_{bt}\text{blk}_4$ removed. The results exemplify that our models achieve comparable stylization performance to the state-of-the-arts in a fraction of the time required by the existing methods. (Part 5/11) }
    \label{fig:stylization5}
\end{figure*}

\begin{figure*}
    \centering
    \includegraphics[width=1.0\textwidth]{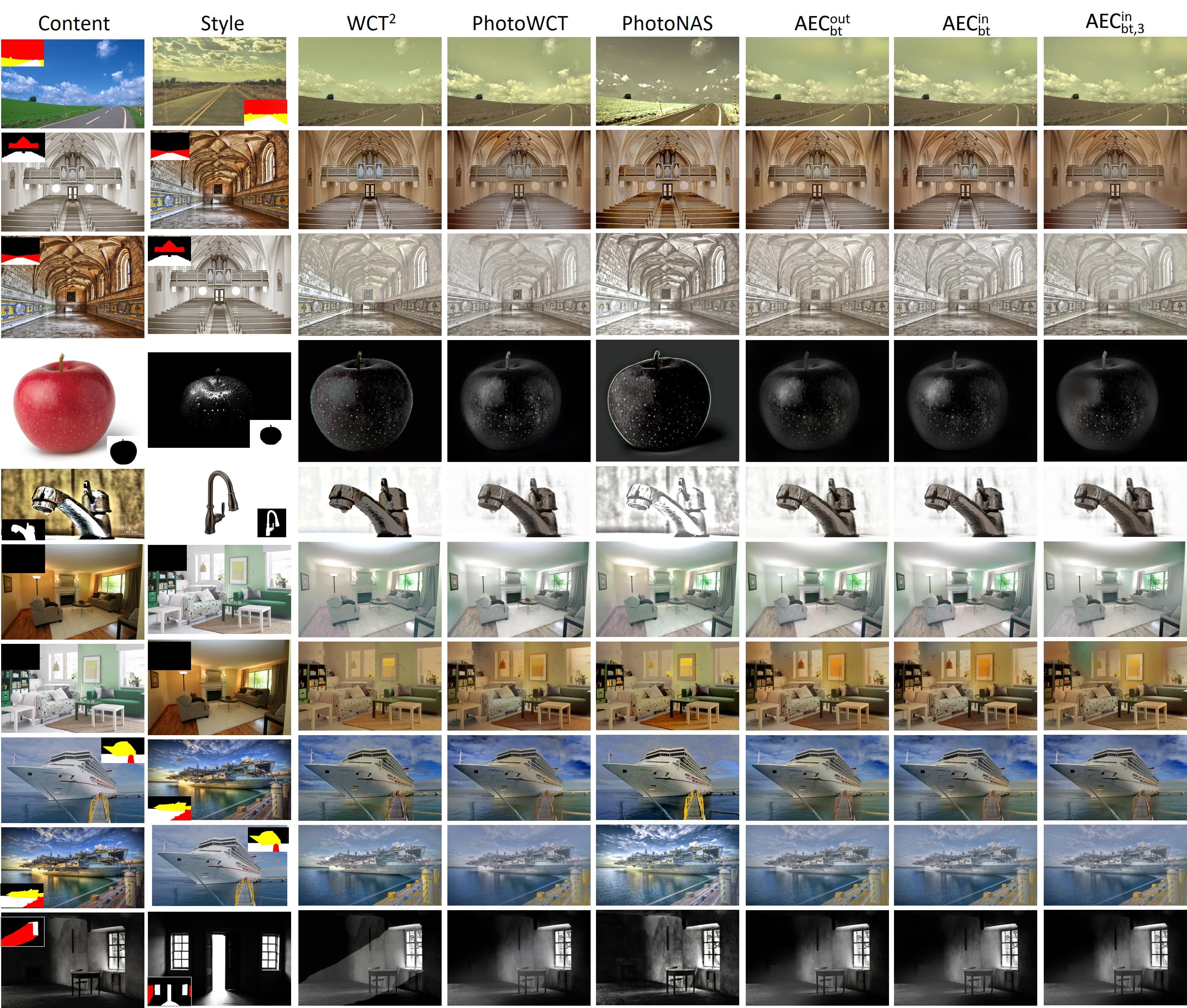}
    \caption{Results of stylization with segmentation for images in the DPST dataset. Three baselines are two state-of-the-arts PhotoWCT~\cite{li2018closed} and WCT$^2$~\cite{yoo2019photorealistic} and a more recent method PhotoNAS~\cite{an2020ultrafast}, while AEC$_{bt}^{in}$ (i.e., PhotoWCT$^2$) and AEC$_{bt}^{out}$ are the autoencoder in Figure~\ref{fig:model} trained inward blockwisely and outward blockwisely, respectively, and AEC$_{bt,3}^{in}$ is AEC$_{bt}^{in}$ with $\text{enc}_4\text{blk}_4$ and $\text{dec}_{bt}\text{blk}_4$ removed. The results exemplify that our models achieve comparable stylization performance to the state-of-the-arts in a fraction of the time required by the existing methods. (Part 6/11) }
    \label{fig:stylization6}
\end{figure*}

\begin{figure*}
    \centering
    \includegraphics[width=1.0\textwidth]{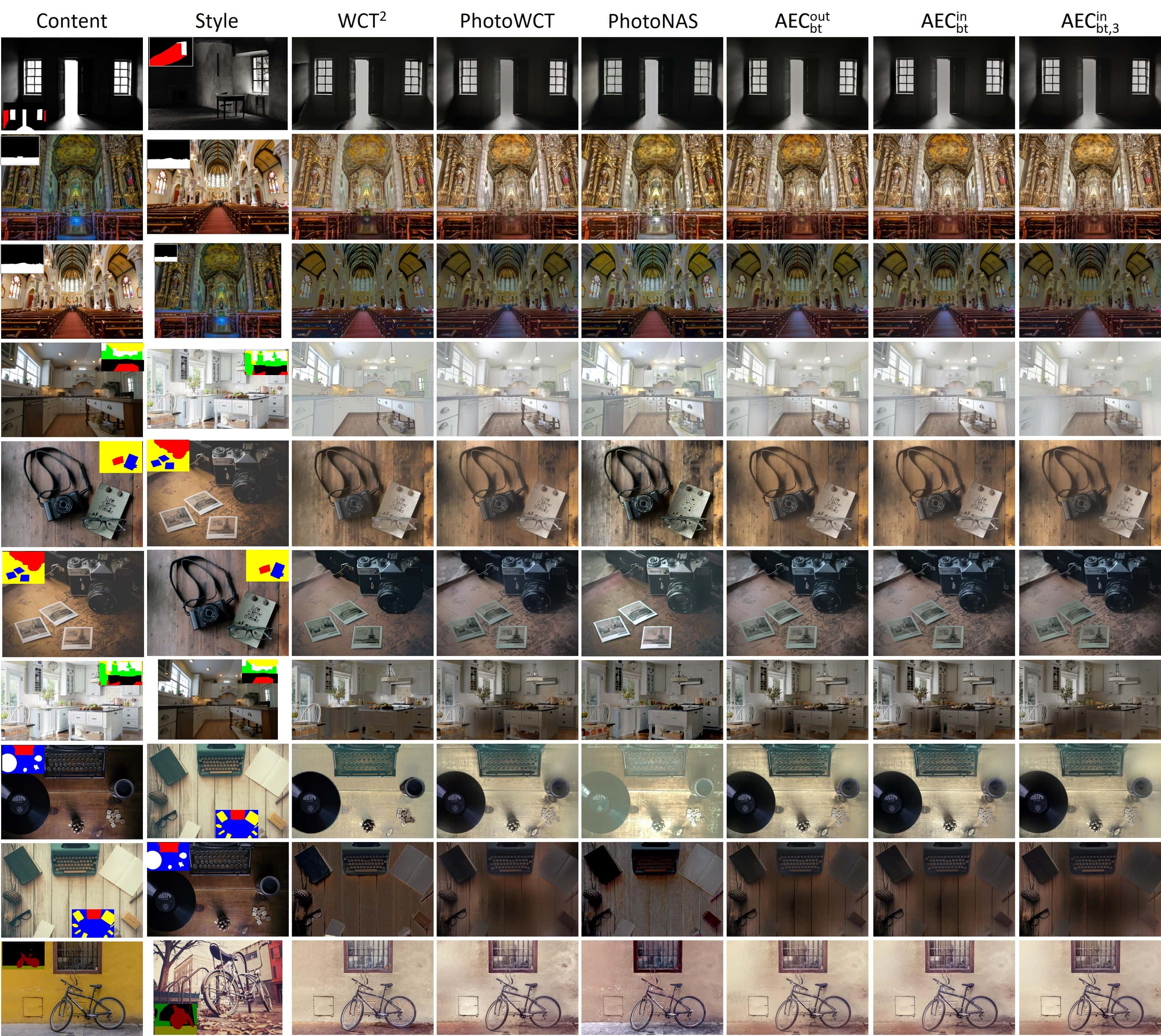}
    \caption{Results of stylization with segmentation for images in the DPST dataset. Three baselines are two state-of-the-arts PhotoWCT~\cite{li2018closed} and WCT$^2$~\cite{yoo2019photorealistic} and a more recent method PhotoNAS~\cite{an2020ultrafast}, while AEC$_{bt}^{in}$ (i.e., PhotoWCT$^2$) and AEC$_{bt}^{out}$ are the autoencoder in Figure~\ref{fig:model} trained inward blockwisely and outward blockwisely, respectively, and AEC$_{bt,3}^{in}$ is AEC$_{bt}^{in}$ with $\text{enc}_4\text{blk}_4$ and $\text{dec}_{bt}\text{blk}_4$ removed. The results exemplify that our models achieve comparable stylization performance to the state-of-the-arts in a fraction of the time required by the existing methods. (Part 7/11) }
    \label{fig:stylization7}
\end{figure*}

\begin{figure*}
    \centering
    \includegraphics[width=1.0\textwidth]{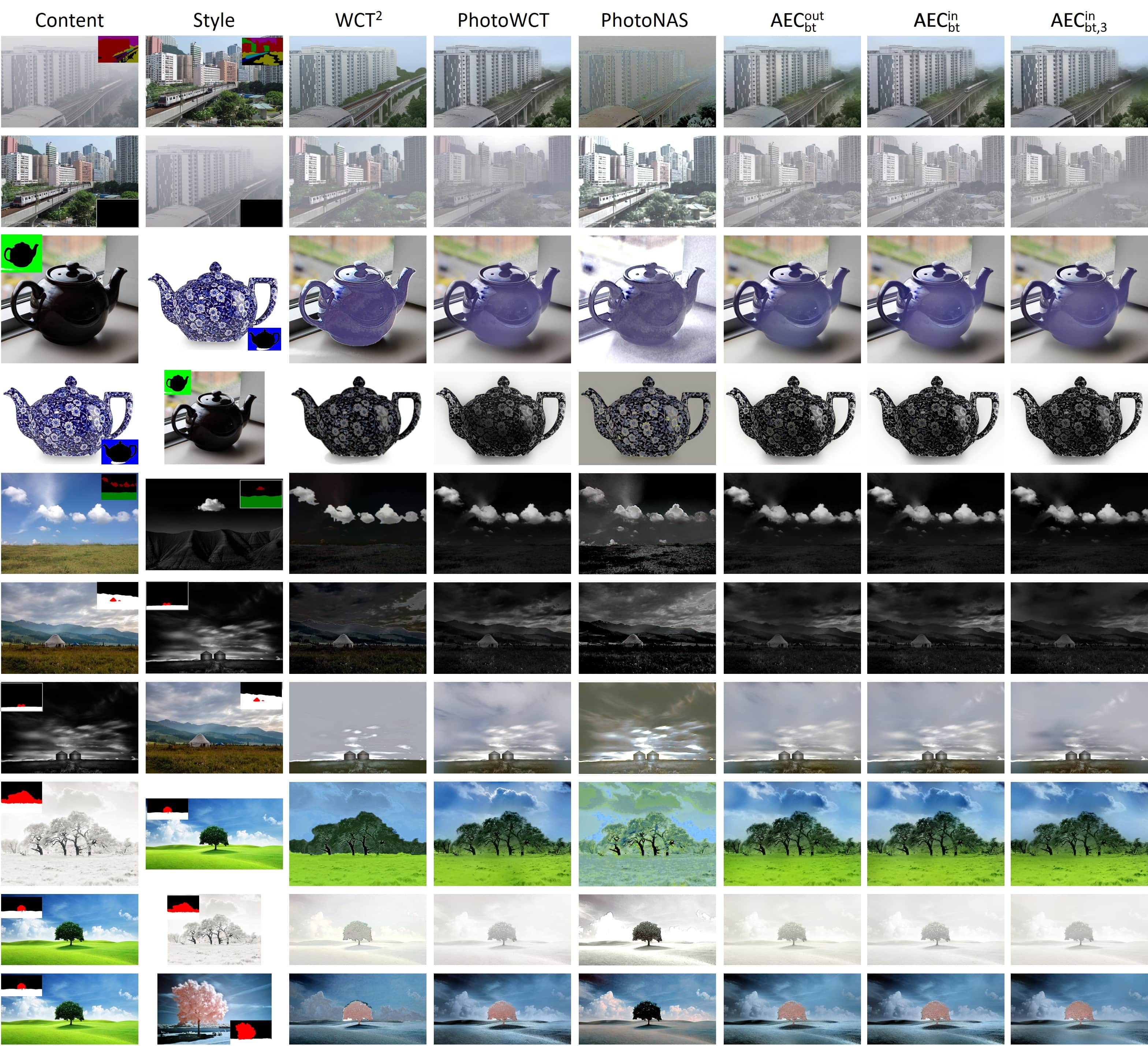}
    \caption{Results of stylization with segmentation for images in the DPST dataset. Three baselines are two state-of-the-arts PhotoWCT~\cite{li2018closed} and WCT$^2$~\cite{yoo2019photorealistic} and a more recent method PhotoNAS~\cite{an2020ultrafast}, while AEC$_{bt}^{in}$ (i.e., PhotoWCT$^2$) and AEC$_{bt}^{out}$ are the autoencoder in Figure~\ref{fig:model} trained inward blockwisely and outward blockwisely, respectively, and AEC$_{bt,3}^{in}$ is AEC$_{bt}^{in}$ with $\text{enc}_4\text{blk}_4$ and $\text{dec}_{bt}\text{blk}_4$ removed. The results exemplify that our models achieve comparable stylization performance to the state-of-the-arts in a fraction of the time required by the existing methods. (Part 8/11) }
    \label{fig:stylization8}
\end{figure*}

\begin{figure*}
    \centering
    \includegraphics[width=1.0\textwidth]{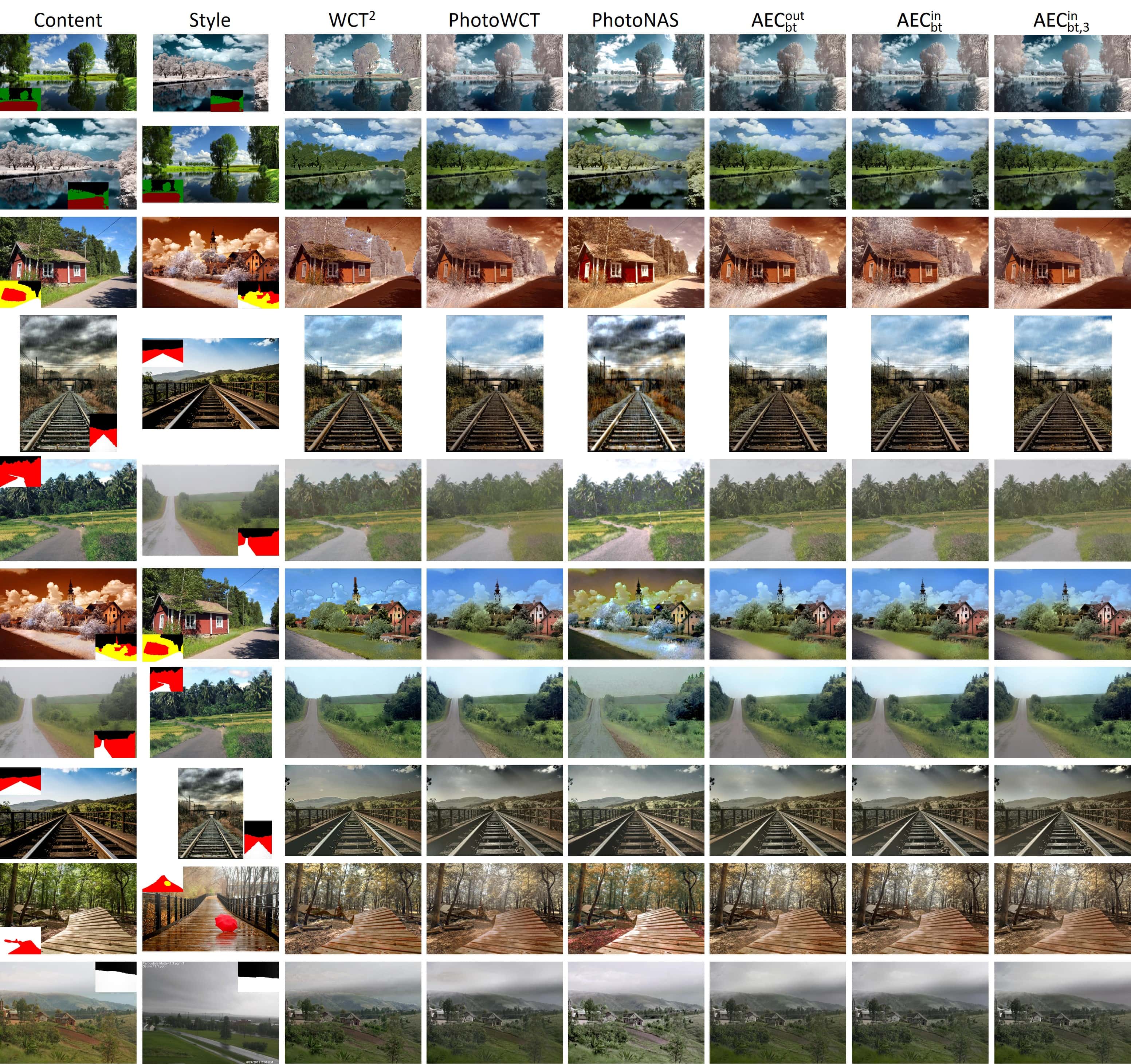}
    \caption{Results of stylization with segmentation for images in the DPST dataset. Three baselines are two state-of-the-arts PhotoWCT~\cite{li2018closed} and WCT$^2$~\cite{yoo2019photorealistic} and a more recent method PhotoNAS~\cite{an2020ultrafast}, while AEC$_{bt}^{in}$ (i.e., PhotoWCT$^2$) and AEC$_{bt}^{out}$ are the autoencoder in Figure~\ref{fig:model} trained inward blockwisely and outward blockwisely, respectively, and AEC$_{bt,3}^{in}$ is AEC$_{bt}^{in}$ with $\text{enc}_4\text{blk}_4$ and $\text{dec}_{bt}\text{blk}_4$ removed. The results exemplify that our models achieve comparable stylization performance to the state-of-the-arts in a fraction of the time required by the existing methods. (Part 9/11) }
    \label{fig:stylization9}
\end{figure*}

\begin{figure*}
    \centering
    \includegraphics[width=1.0\textwidth]{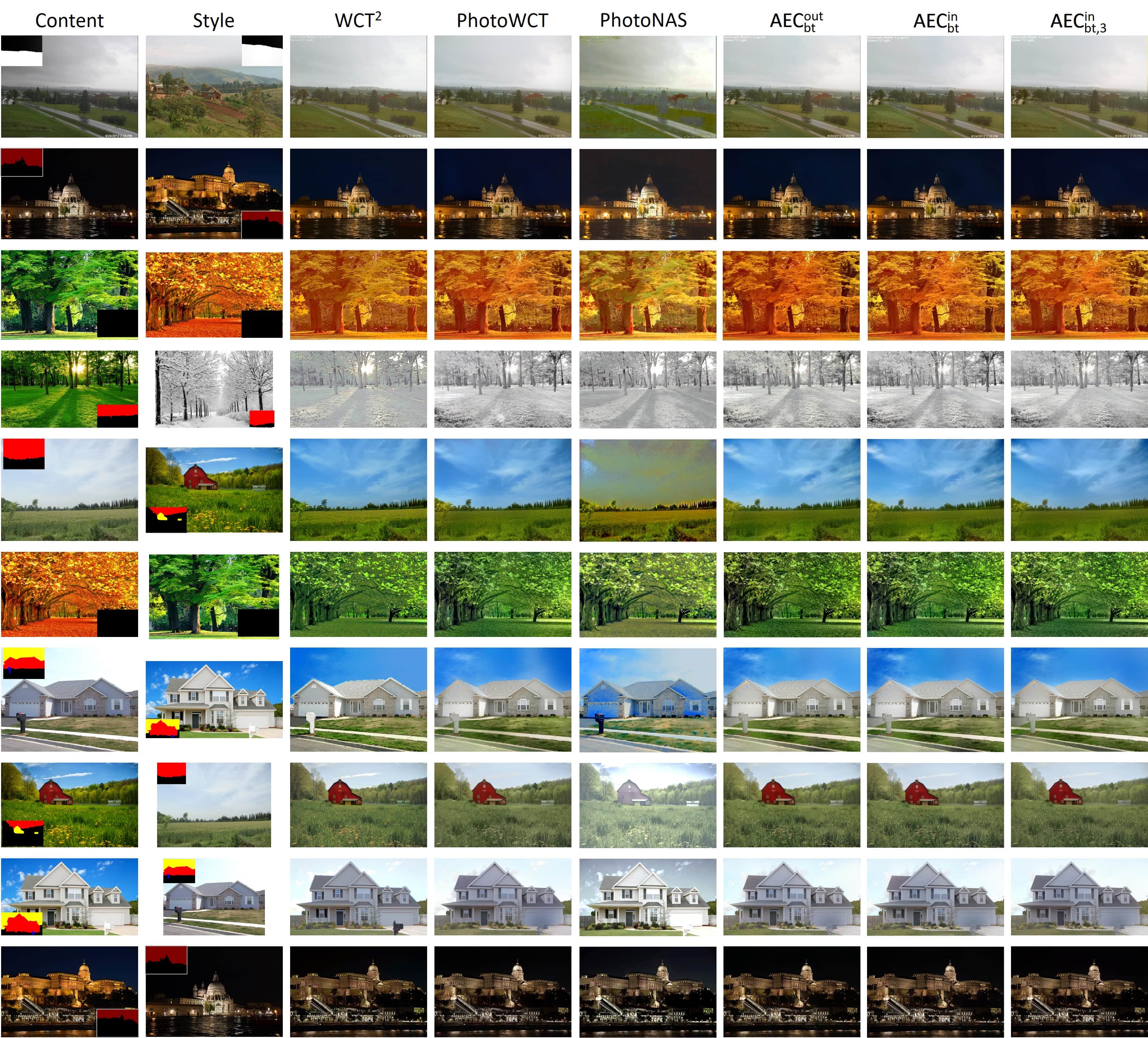}
    \caption{Results of stylization with segmentation for images in the DPST dataset. Three baselines are two state-of-the-arts PhotoWCT~\cite{li2018closed} and WCT$^2$~\cite{yoo2019photorealistic} and a more recent method PhotoNAS~\cite{an2020ultrafast}, while AEC$_{bt}^{in}$ (i.e., PhotoWCT$^2$) and AEC$_{bt}^{out}$ are the autoencoder in Figure~\ref{fig:model} trained inward blockwisely and outward blockwisely, respectively, and AEC$_{bt,3}^{in}$ is AEC$_{bt}^{in}$ with $\text{enc}_4\text{blk}_4$ and $\text{dec}_{bt}\text{blk}_4$ removed. The results exemplify that our models achieve comparable stylization performance to the state-of-the-arts in a fraction of the time required by the existing methods. (Part 10/11) }
    \label{fig:stylization10}
\end{figure*}

\begin{figure*}
    \centering
    \includegraphics[width=1.0\textwidth]{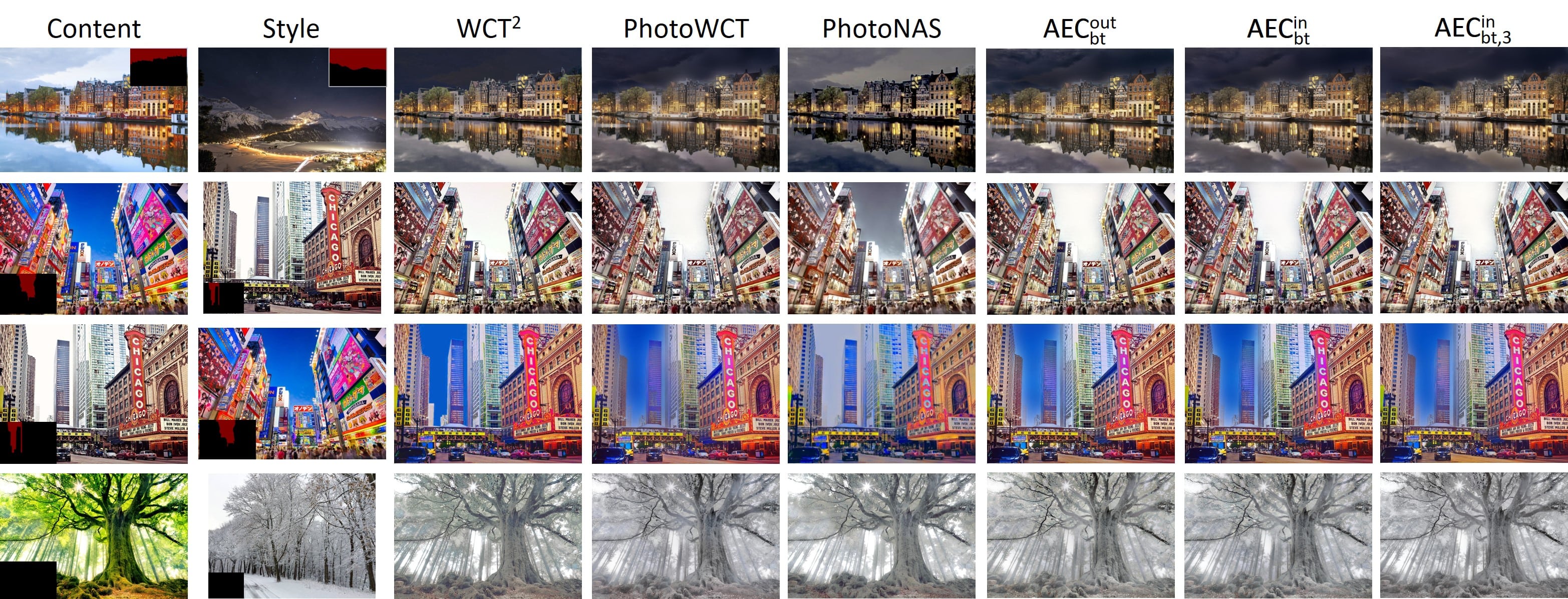}
    \caption{Results of stylization with segmentation for images in the DPST dataset. Three baselines are two state-of-the-arts PhotoWCT~\cite{li2018closed} and WCT$^2$~\cite{yoo2019photorealistic} and a more recent method PhotoNAS~\cite{an2020ultrafast}, while AEC$_{bt}^{in}$ (i.e., PhotoWCT$^2$) and AEC$_{bt}^{out}$ are the autoencoder in Figure~\ref{fig:model} trained inward blockwisely and outward blockwisely, respectively, and AEC$_{bt,3}^{in}$ is AEC$_{bt}^{in}$ with $\text{enc}_4\text{blk}_4$ and $\text{dec}_{bt}\text{blk}_4$ removed. The results exemplify that our models achieve comparable stylization performance to the state-of-the-arts in a fraction of the time required by the existing methods. (Part 11/11) }
    \label{fig:stylization11}
\end{figure*}

We show the stylization results for the DPST dataset in Figures~\ref{fig:stylization1} to \ref{fig:stylization11}. Each row in the figures contains the results of photorealistic style transfer methods from a pair of content and style images and the associated segmentation. In particular, a segment in a content image is rendered with the style of the corresponding segment denoted in the same color in the style image. Note that the results from PhotoNAS~\cite{an2020ultrafast} do not use the segmentation labels, since the PhotoNAS paper clearly states ``the proposed algorithm allows transferring photo styles without any assist of region masks acquired by segmenting content and style inputs" and the official code does not support this feature, either.

\clearpage

\clearpage

\clearpage

\clearpage

\clearpage

\clearpage

\clearpage

\clearpage

\clearpage

\clearpage

\clearpage

{\small
\bibliographystyle{ieee_fullname}
\bibliography{egbib}
}

\end{document}